\documentclass[12pt,titlepage]{utarticle-dan}

\pdfoutput=1

\usepackage{amsmath,amsthm,amsfonts,amscd,amssymb} 
\usepackage{eucal}
\usepackage{braket}
\usepackage{slashed}
\usepackage{hyperref}
\usepackage{graphicx}
\usepackage{bm}
\usepackage{siunitx}
\usepackage{xcolor}
\usepackage{afterpage}
\usepackage{changepage}
\usepackage{cite}

\usepackage{authblk}
\newcommand{\Oi}{\mathcal{O}}

\newcommand{\be}{\begin{equation}}
\newcommand{\ee}{\end{equation}} 
\newcommand{\mb}{\mathbf}

\newcommand{\bear}{\begin{eqnarray}}
\newcommand{\eear}{\end{eqnarray}}

\begin{document}

\title{Tabletop experiments for quantum gravity: \newline \newline a user's manual}
\author[1,2]{Daniel Carney\footnote{Electronic address: carney@umd.edu}}
\author[3,4,5]{Philip C. E. Stamp}
\author[1,2]{Jacob M. Taylor}
\affil[1]{Joint Quantum Institute, National Institute of Standards and Technology \newline Gaithersburg, Maryland 20899, USA}
\affil[2]{Joint Center for Quantum Information and Computer Science, University of Maryland \newline College Park, Maryland 20742, USA}
\affil[3]{Pacific Institute of Theoretical Physics, University of British Columbia \newline Vancouver, B.C. V6T 1Z1, Canada}
\affil[4]{Department of Physics and Astronomy, University of British Columbia \newline Vancouver, B.C. V6T 1Z1, Canada}
\affil[5]{School of Mathematics and Statistics, Victoria University of Wellington \newline Wellington 6140, New Zealand}

\Abstract{Recent advances in cooling, control, and measurement of mechanical systems in the quantum regime have opened the possibility of the first direct observation of quantum gravity, at scales achievable in experiments. This paper gives a broad overview of this idea, using some matter-wave and optomechanical systems to illustrate the predictions of a variety of models of low-energy quantum gravity. We first review the treatment of perturbatively quantized general relativity as an effective quantum field theory, and consider the particular challenges of observing quantum effects in this framework. We then move on to a variety of alternative models, such as those in which gravity is classical, emergent, or responsible for a breakdown of quantum mechanics.}

\maketitle

\newpage

\tableofcontents

\section{Introduction}

The dream of measuring gravitational effects in the quantum regime dates to the earliest days of general relativity \cite{einstein1916naherungsweise,rosenfeld1930quantelung,bronstein1936quantentheorie}. Experiments have shown that classical gravitational fields, such as that of the Earth or a gravitational wave, act on quantum systems in the same manner as other external potentials, allowing for neutron guiding, cold atom trapping, and even displacement of interferometric mirrors \cite{colella1975observation,nesvizhevsky2002quantum,abbott2016observation}. However, observation of the gravitational field produced by a mass prepared in a distinctly quantum state has yet to be experimentally achieved. This concept has driven significant physical and philosophical interest, and experimental proposals have begun to appear \cite{bose2017spin,marletto2017entanglement} based on earlier suggestions \cite{cecile2011role,feynman1971lectures,page1981indirect}.

Recent, rapid progress in the quantum control of meso-to-macroscopic mechanical systems suggests that such experiments may be feasible in the near future. For example, the development of long-lifetime mechanical systems has enabled a dramatic push towards realization of the quantum ground state for the center-of-mass motion of objects with masses ranging up to the kilogram scale\cite{king1998cooling,LaHaye2004,Marquardt2007,corbitt2007all,Cleland2010,teufel,TheLIGOScientific:2014jea}. Beyond preparing the ground state, there are a variety of approaches for creating and monitoring non-classical excited states of macroscopic mechanical oscillators. Most methods couple the mechanical devices to electromagnetism in the optical and microwave domains, and use squeezed light \cite{walls1983squeezed,wu1986generation,Wineland1994,Clerk2008,Purdy2013}, photon number eigenstates \cite{cirac1993preparation,bertet2002direct,hong2017hanbury,ringbauer2018generation,clarke2018growing}, or more recently the stabilization of Schr\"{o}dinger-cat type states via measurement\cite{gottesman2001encoding,vlastakis2013deterministically}. In a complementary approach, matter-wave interferometry has already demonstrated coherent spatial superpositions of masses around $10^5$ amu with spatial separation on the order of microns \cite{gerlich2011quantum,eibenberger2013matter,romero2017coherent}, and proposals up to the nanogram scale exist \cite{brand2017matter,pino2018chip}. These two types of systems could potentially be used as sources to prepare a measurable, non-classical state of the gravitational field. The advent of this experimental situation demands precise thinking about these issues: we must take concrete models of gravity and make definite predictions for these experiments.

Thus, in this paper, we give an overview of the basic mechanics and predictions of a number of theories of low-energy gravity in some simple, paradigmatic thought experiments. We focus our efforts on experiments involving two types of source masses: freely moving massive test particles in superposition, corresponding to an idealized interferometer, and mechanical resonators prepared in  non-classical states. We study three categories of gravitational models: gravity as a quantum interaction, as a classical interaction, and models in which gravity is somehow responsible for a breakdown of quantum mechanics. We argue that a number of these models, including the simplest model of quantum gravity, may be observable with near-future experimental techniques. 

We begin by explaining how to view perturbative general relativity as a quantum theory. This has historically been viewed as problematic due to the non-renormalizable nature of the theory, but in modern language, this simply means that we are dealing with gravity as an \emph{effective} quantum field theory. In this sense, we have a perfectly good quantum theory of gravity, which gives precise predictions in the kind of experimental circumstances discussed in this paper. We argue that the most promising avenue is to look for gravitationally-generated entanglement coming from the Newtonian interaction between two massive objects; we also study effects involving gravitons and explain why these will be much harder to detect.

As a foil to the quantum model, we then study the idea that gravity could be ``fundamentally classical''. This could potentially mean a number of things, so we begin by discussing the semiclassical or ``Schr\"{o}dinger-Newton'' model in which quantum matter is coupled to a classical gravitational field through expectation values. This model is known to have fundamental theoretical inconsistences because it amounts to a non-linear modification of the Schr\"{o}dinger equation. In order to circumvent these difficulties, we show how to obtain the semiclassical interaction from a self-consistent, unitary quantum model based on measurement and feedback, and discuss how these classical models differ in their predictions from the quantum model.

Beyond these models of the gravitational interaction, a number of authors have suggested various ways in which quantum mechanics might suffer some kind of breakdown due to gravitational effects. We give particular attention to models like those of Penrose and Diosi in which gravitational effects cause the collapse of wavefunctions of macroscopic superpositions. These are non-relativistic models, and so we finish with a discussion of a relativistic variant, ``correlated worldline theory'', involving the breakdown of the usual superposition principle via extra gravitationally-induced correlations, which are capable of explaining the quantum-to-classical transition.

We have endeavored to give a representative selection of topics and references. In particular, we are focusing here on explicit models of the gravitational interaction, as opposed to more phenomenological effects sometimes posited as low-energy consequences of quantum gravity, eg. \cite{Pikovski2012}. Much work has been done on many aspects of what we will cover and while we have attempted to give a detailed list of citations in the text, our primary goal is to collate a user-friendly overview. For more exhaustive treatments of some background on topics influencing this work, we refer the interested reader to reviews of optomechanics \cite{Chen:2013sgs,RevModPhys.86.1391}, electromechanics \cite{blencowe2004quantum}, matter-wave interferometry \cite{arndt2014matter}, quantum sensing \cite{RevModPhys.89.035002,clerk2010introduction}, and gravitational decoherence \cite{bassi2017gravitational}.

This paper is organized as follows. In section \ref{simplemodels} we give a brief overview of our gedanken matter-wave and resonator systems, and provide a summary table of the behavior and predictions of the models studied in the paper. We then move on in section \ref{EFTsection} to the study of general relativity as an effective quantum field theory. Classical models of gravity are studied in section \ref{classical-gravity}, and models of gravitationally-induced breakdown of quantum mechanics are discussed in section \ref{alt-grav-models}. We end with our conclusions in section \ref{conclusions}.

\section{Some model systems and experiments}
\label{simplemodels}

\begin{figure}[t]
\begin{center}
\includegraphics[scale=0.9]{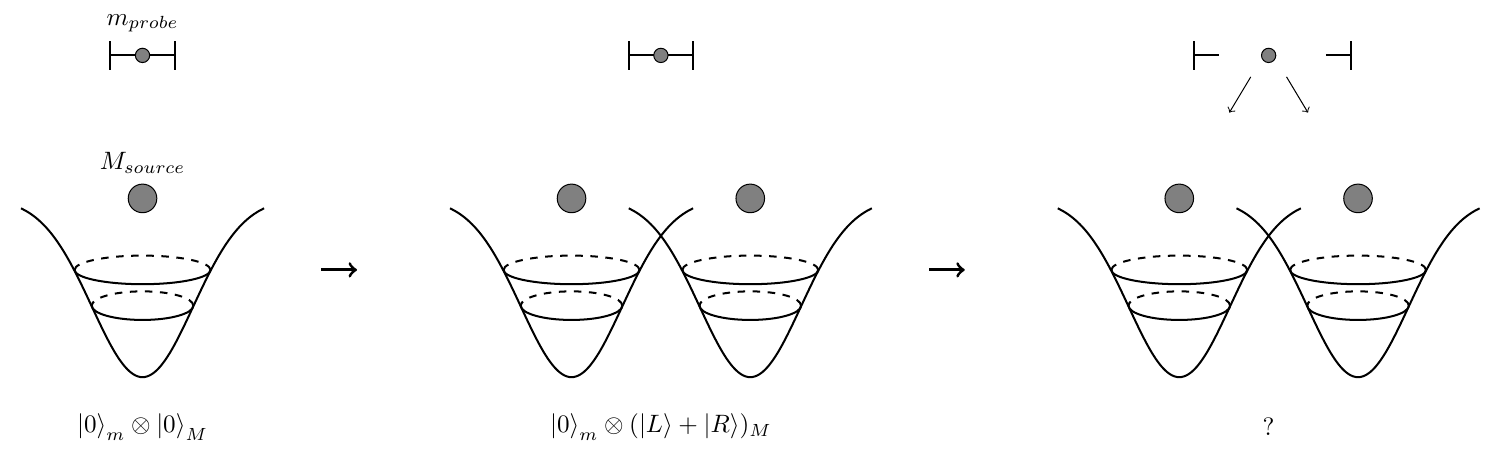}
\end{center}
\caption{A cartoon of our experimental paradigm. In the first step, a source mass $M$ is prepared in some initial state like the ground state $\ket{0}_M$, which produces a particular gravitational field, here depicted as a small gravitational potential well, while a test mass $m$ is likewise prepared in some reference state $\ket{0}_m$. We then prepare the source into a non-classical state, here a cat state $\ket{L} + \ket{R}$ of two locations, each branch of which has its associated gravitational field. Finally, we let the test mass $m$ freely interact with the source mass, and perform a measurement on either one of the masses or the joint source-test system.}
\label{schematic}
\end{figure}

The essential style of experiment we imagine consists of preparing a massive object in a non-classical state, such as a superposition of two locations. We then consider different possible probes of the gravitational response to this source. Typically we consider a nearby test mass responding to the resulting state-dependent force. See figure \ref{schematic} for a schematic of this process. As an intermediate experiment, we can simply try to prepare the source mass in a non-classical state and measure its coherence time; this can be used to rule out models of gravitational decoherence.

\begin{figure}
\centering
\includegraphics[scale=0.8]{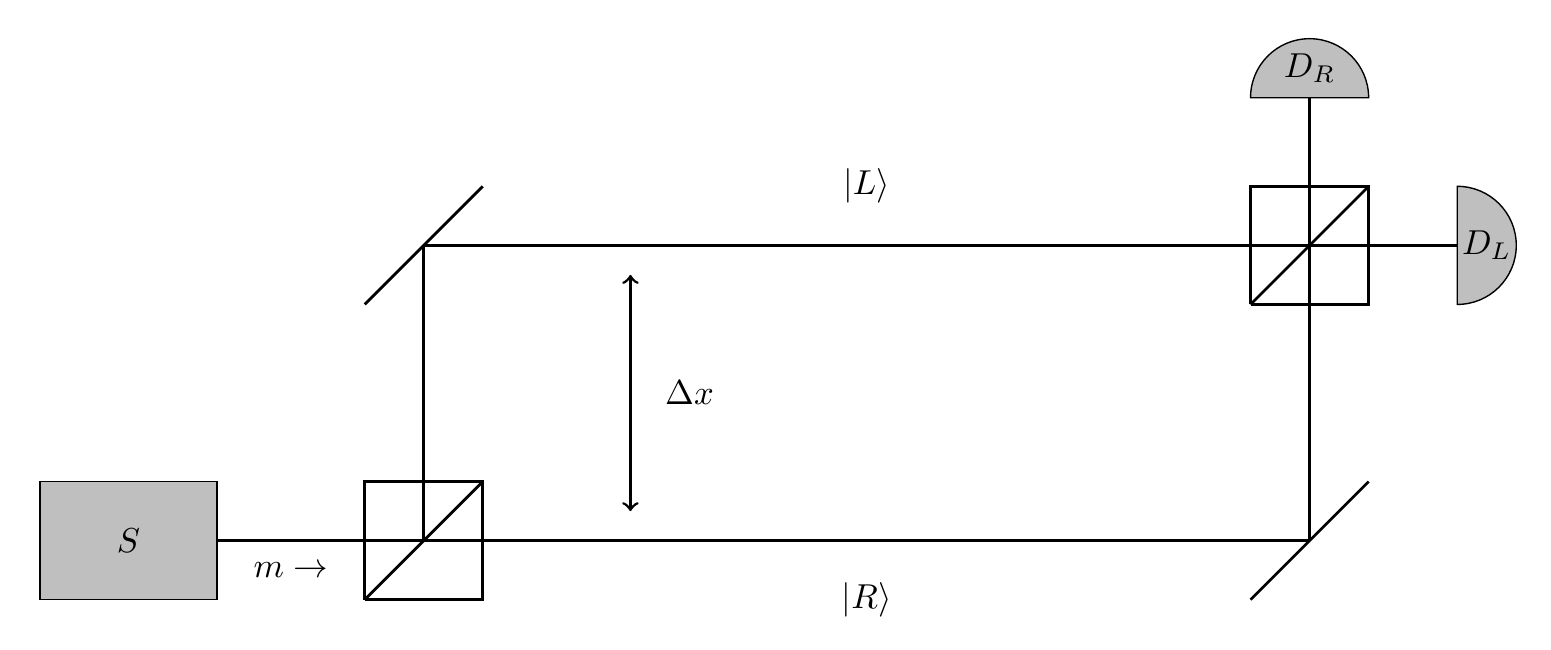}
\caption{Schematic of a (Mach-Zehnder) matter-wave interferometry experiment. A beam of massive particles is sent from the source $S$ through a beamsplitter, evolves freely along the beam arms, and then recombined and counted in the detectors $D_{L,R}$. The observation of this coherent superposition of the massive beam already probes gravitationally-induced decoherence; one could further imagine looking for gravitational interactions between this cat state and a nearby test particle.}
\label{cartoon-mw} 

\vspace{1cm}

\includegraphics[scale=.8]{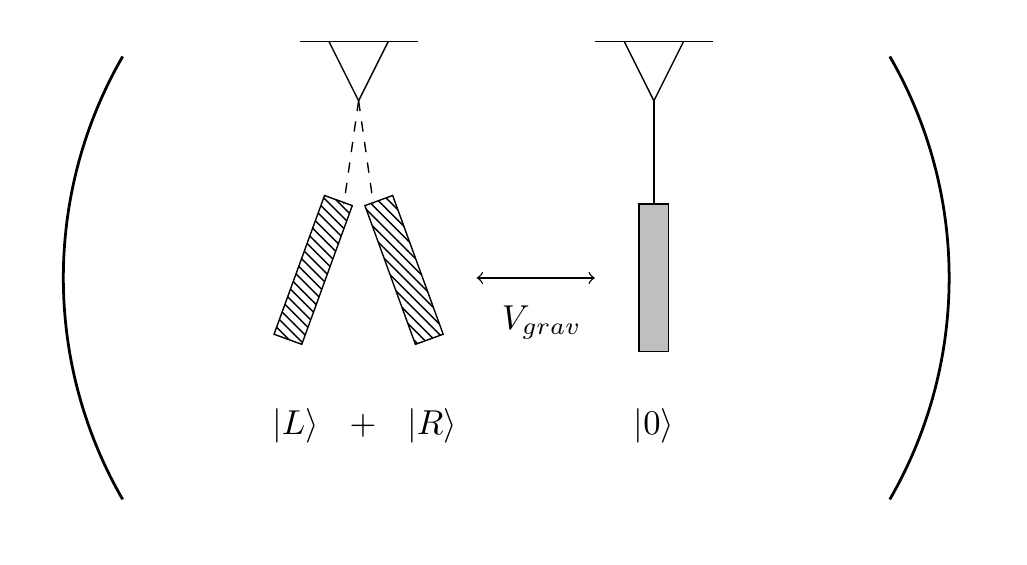}
\caption{Schematic of a cavity optomechanics experiment involving two suspended mirrors. Here, one mirror is depicted in a ``cat'' state, a superposition of two coherent states, and viewed as a source mass. The other mirror is then prepared in its ground state and used as a gravitational detector. The goal of an experiment with such a setup would be to try to see the superposed source mass entangle with the test mass; see section \ref{eft-newton}.}
\label{cartoon-opto}
\end{figure}

Consider first a matter-wave interferometry experiment. We prepare a cold, massive particle and send this through a beam-splitter, producing a state of the form $\ket{L} + \ket{R}$, corresponding to two distinct paths with spatial separation $\Delta \mb{x}$. The particle is then allowed to freely propagate for some time $\Delta t$, before we recombine it with an inverse beam splitter and projectively measure the resulting state; see figure \ref{cartoon-mw}. 

The state of the mass as it propagates down the interferometer arms in superposition is an example of the kind of Schr\"{o}dinger-cat type state we are interested in. Simply observing the coherent interference of this beam after the free propagation sets bounds on any model of gravitational decoherence. Furthermore, we could use the superposed beam as a gravitational source mass, and try to observe its coupling to some other nearby test mass system. For example, the usual quantum treatment of the gravitational interaction (see section \ref{EFTsection}) predicts entanglement generation between the source and test mass; other models (sections \ref{classical-gravity}, \ref{alt-grav-models}) will predict other behaviors. 

A complementary approach is to use a mechanical resonator system. Here, the masses do not propagate freely but rather oscillate about fixed locations via a restoring force from a spring or other harmonic confinement mechanism. A prototypical example is a high-reflectivity mirror suspended in an optical cavity. The mirror's spatial position can be both controlled and measured by cavity photons. Other examples include optically trapped particles, mechanical cantilevers, and high-tension membanes.

In analogy with the matter-wave experiment above, we might imagine cooling the center-of-mass motion of the resonator to its ground state $\ket{0}$, and then preparing some non-classical state, for example the cat-type superposition $\ket{\alpha} + \ket{-\alpha}$ of two coherent states of the center-of-mass motion. Just like the interferometric cat state, this superposition will be sensitive to any gravitational decoherence effects, and similarly, could be used as a gravitational source.

We note that the roles of these two types of massive objects--and indeed other, similar systems--can be mixed and matched. An experiment could use a resonator as a source mass and an interferometer as a test mass, or vice versa. A key difference between these two systems is the vastly different timescales probed: matter-wave interferometers have bandwidths set by their time of flight, whereas resonator systems oscillate with their mechanical frequency.

For ease of use, in figure \ref{tablesummary} we provide a table summarizing the models studied in the rest of this paper. We give a brief description of some basic properties of the model, and give an example prediction for both a single-body coherence experiment and two-body entanglement experiment of the type described above.

\begin{figure}
 \begin{adjustwidth}{-1.6cm}{}

\begin{center}
\small
\begin{tabular}{| m{2.6cm} | m{2.6cm} | m{2.6cm} | m{2.6cm} | m{2.6cm} | m{2.6cm} |}
\hline
 & EFT & Semi-classical/SN* & Classical channel & Penrose-Diosi & Correlated worldline \\
 \hline
 \hline
Entanglement? & Yes & No & No & ** & Yes \\
 \hline
 Decoherence mechanisms & Graviton emission; gravitational coupling to unobserved matter & None & Anomalous heating & Gravitational ``noise'' & Graviton emission; gravitational coupling to unobserved matter \\
 \hline
 QM modifications & None & Non-linear Schr\"{o}dinger equation & None & Wavefunction collapse for large superpositions & Breakdown of superposition principle, extra gravitational correlations \\
 \hline
 \hline
 \multicolumn{6}{|c|}{\bf{Single-body/quantum coherence experiments}} \\
 \hline
 \hline
Signal & Graviton-induced decoherence & Wavefunction self-attraction & Anomalous heating & Anomalous gravitational decoherence & Path bunching \\
\hline
Falsified if & Coherence time $> \Delta t$ & Coherence with masses $m > m_{SN}$ & Coherence time $> \Delta t$ & Coherence time $> \Delta t$ & Coherence with masses $m > m_{CWL}$ \\
\hline
Example & Resonator decoherence from spontaneous emission: $\Delta t = {\hbar c^5}/{G_N Q^2 \omega^5}$ & Talbot-Lau interferometry: $m_{SN} = (\hbar^2/G_N \sigma_x)^{1/3}$ & BEC interferometry: $\Delta t = k_B T_{c} R_0^3/G_N \hbar m$ & Matter-wave inteferometry $\Delta t = {\hbar R_0}/{G_N m^2}$ &  \\

\hline
\hline
 \multicolumn{6}{|c|}{\bf{Two-body/entanglement experiments}} \\
 \hline
 \hline
Signal & Entanglement of positions & No entanglement & No entanglement & ** & Entanglement of positions, only for masses $m < m_{CWL}$ \\
\hline
Falsified if & No entanglement generated & Entanglement generated & Entanglement generated & ** & Entanglement generated with $m > m_{CWL}$ \\
\hline
Example & Two resonators, interference terms controlled by phase $\Delta \phi = {G_N m \Delta t}/{\hbar \omega d^3}$ & & & ** &   \\
\hline
 \end{tabular}
\end{center}
\label{tablesummary}
\caption{Summary of some properties and example predictions of models studied in this work. See the relevant sections for details and references. Variables used, from left to right: $Q =$ quadrupole moment, $\sigma_x =$ initial position uncertainty in a wavepacket, $T_c =$ BEC critical temperature, $R_0 = $ fundamental length scale in a model, $m = $ single-particle mass. * ``Predictions'' here mean with naive interpretation of the Born rule, see section \ref{alt-grav-models} for a discussion of this point. ** Gravitational dynamics other than collapse mechanism unspecified in this model.}
 \end{adjustwidth}
\end{figure}

\newpage

\section{General relativity as an effective quantum field theory}
\label{EFTsection}

At the terrestrial energies we are concerned with in this paper, general relativity can be treated quantum mechanically in a straightforward manner. It is sometimes said that general relativity and quantum mechanics are mutually inconsistent, but this is only really true at extremely high energies, where the non-renormalizable nature of the theory becomes problematic; in this limit, untameable divergences arise. The key is to realize that perturbatively quantized general relativity is an \emph{effective} quantum field theory (EFT), valid only below some very high energy scale, in this case the Planck scale $M_{p} c^2 \sim \SI{e19}{\giga\electronvolt}$ \cite{Weinberg:1978kz,Georgi:1994qn,Donoghue:1994dn,Donoghue:1995cz,Burgess:2003jk}. 

In the kinds of low-energy settings we have in mind, the EFT of general relativity is easy to summarize. The gravitational field $g_{\mu\nu}$ is expanded around a fixed classical background, say flat spacetime $\eta_{\mu\nu}$, as 
\be
g_{\mu\nu} = \eta_{\mu\nu} + h_{\mu\nu},
\ee
and the small fluctuations $h_{\mu\nu} = h_{\mu\nu}(\mb{x},t)$ are a quantized field. Note that $h_{\mu\nu}$ contains both a longitudinal ``Newtonian'' component as well as dynamic ``graviton'' fluctuations. In particular, the gravitational field can transmit quantum information and generate entanglement. If one deals with large-amplitude deviations from the reference classical field, the effective description breaks down, but the experiments discussed in this paper are safely within the limits of the EFT treatment. 

In the non-relativistic limit, massive objects interact via the usual $1/r$ Newton potential, treated as a quantum operator on the Hilbert space of the masses. Superpositions of states of massive objects lead to superpositions of the metric, which in the non-relativistic limit means superpositions of the Newton potential, as depicted in figure \ref{schematic}. Additionally, the transverse, dynamical metric fluctuations themselves are massless spin-2 particles, called gravitons, which couple to any source of energy including rest mass. Structurally, the theory is very similar to quantum electrodynamics with its Coulomb potential and photons, and one can typically import their intuition from QED to gravity at these energies. The non-linearity of the gravitational interaction is negligible, since gravity couples to energy and the energy stored in the gravitational fields here is tiny. 

The EFT paradigm organizes predictions as a Taylor series in some small dimensionless parameters, for example the ratio 
\be
\label{taylorseries}
\epsilon = \frac{E}{M_p c^2}
\ee
of the energy transfer $E$ during a process in units of the Planck energy. Each term in the Taylor series comes with a free, unknown coefficient which in principle needs to be fixed by experiments. The non-renormalizability of the model means that whenever we are dealing with processes where $E/M_p c^2$ becomes a sizable fraction of unity, our ability to make predictions breaks down, since all the terms in the series become important. For processes where $E/M_p c^2$ is small, on the other hand, we can ignore these higher-order terms.

As a concrete example of this approach, one can explicitly compute the leading quantum corrections to the Newtonian potential \cite{Burgess:2003jk,BjerrumBohr:2002kt}. Feynman diagrams with a single graviton loop give
\be
\label{eftpot}
V(r) = - \frac{G_N m_1 m_2}{r} \left( 1 + \lambda \frac{G_N (m_1 + m_2)}{c^2 r} + \xi \frac{G_N \hbar}{c^3 r^2} +\cdots \right)
\ee
where the Newton constant $G_N \approx \SI{6e-11}{\newton \meter^2 \kilogram^{-2}}$, and the constants $\lambda,\xi$ are calculable, $\Oi(1)$ quantities whose precise values depend on exactly how one defines the potential. The term proportional to $\lambda$ is the first ``post-Newtonian'' correction, which is classical, while the term proportional to $\xi$ is the first quantum correction. The ellipses represent terms of higher order in the Newton constant. With very conservative estimates of $m \sim \SI{1}{\milli\gram}$ and $r \sim \SI{1}{\micro\meter}$, the displayed corrections are of order $10^{-28}$ and $10^{-58}$, respectively, and are utterly negligible.

It should be emphasized that the effective field theory picture given here is the universal low-energy limit of any theory which contains massless spin-$2$ excitations and has an $S$-matrix \cite{Weinberg:1964ew,Weinberg:1965rz,deser1970self,boulware1975classical}. For example, in string theory, each closed string has a massless spin-$2$ state. Thus a detection of some violation from the EFT predictions would give direct information about the ultraviolet nature of quantum gravity. 

There is one important subtlety, which is how to properly treat the composite objects used in these experiments as particles with rest mass approaching or greater than the Planck mass $M_{p} \sim \SI{e-5}{\gram}$. One might worry about $N$-body effects; for example, a relic graviton in the cosmic microwave background has temperature of about $\SI{1}{\kelvin}$ \cite{kolb2018early} and thus a wavelength of about a millimeter, so it could probe internal structure of some of these objects. In a proper EFT treatment this will need to be taken into account through form factors encoding the details of the massive objects. One may also worry about the validity of a Taylor series like \eqref{taylorseries} for objects with $M \gtrsim M_p$. Although the description given above will provide the leading order effects, a detailed treatment along the lines of heavy quark EFT \cite{Isgur:1989vq,Georgi:1990um} or the chiral theory of nucleons \cite{Weinberg:1978kz}--effective models involving rest masses significant larger than the QCD scale--will be crucial for understanding subleading effects. See for example \cite{Goldberger:2004jt,Goldberger:2005cd} for some classical EFT treatments in gravity along these lines.

\subsection{Tests of the quantum Newtonian interaction}
\label{eft-newton}

The most basic prediction of the EFT picture is that massive objects will source Newtonian gravitational fields tied to their center-of-mass positions. If we prepare a source mass in a non-classical state and bring in a test mass, gravity can entangle the two masses, and it is this effect that we could like to witness. We begin by explaining this idea using resonator systems, and then review the matter-wave variant proposed in \cite{bose2017spin,marletto2017entanglement}.

Consider a pair of harmonic oscillators of equal mass $m$ and frequency $\omega$. For simplicity we will completely ignore any non-gravitational interaction between the objects. In practice, methods for distinguishing the gravitational from non-gravitational interactions will be needed. We also ignore any mechanical dissipation here. We defer these issues to more detailed work; the goal here is just to work through the simple points of principle. Thus the Hamiltonian reads
\be
H = \sum_{i=1,2} \frac{\mb{p}_i^2}{2m} + \frac{1}{2} m \omega^2 \mb{x}_i^2 - \frac{G_N m^2}{|\mb{d} -(\mb{x}_1 - \mb{x}_2)|}.
\ee
Here $\mb{d}$ is the vector between the two equilibrium positions of the oscillators. Assuming that $|\mb{d}| \gg |\mb{x}_1 - \mb{x}_2|$, we can expand out the denominator in a multipole expansion. The zeroth order term is an overall constant and may be dropped, the first order term is proportional to $\mb{x}_1 - \mb{x}_2$ and can thus be eliminated by re-definition (at order $G_N$) of the equilibrium positions, and the second-order term contains both $\mb{x}_1^2 + \mb{x}_2^2$ and $\mb{x}_1 \cdot \mb{x}_2$. The squared terms can similarly be eliminated by re-definition (at order $G_N$) of the oscillator frequencies. The order $G_N$ corrections are negligible and we drop them. For further simplicity we will restrict to a single spatial dimension, by eg. appropriately designing the oscillators or sharply trapping the particles in the transverse directions. Finally, we take the rotating wave approximation, which here is excellent due to the tiny coupling, allowing us to drop the interaction terms $a_1 a_2 + a_1^{\dagger} a_2^{\dagger}$. In the end, we obtain the Hamiltonian
\be
\label{2osc-H}
H = \sum_{i=1,2} \hbar \omega a_i^{\dagger} a_i - \hbar \lambda_g (a_1 a_2^{\dagger} + a_1^{\dagger} a_2) 
\ee
where the coupling constant is
\be
\label{osc-coupling}
\lambda_g = \frac{G_N m^2 x_{ZPF}^2}{\hbar d^3} = \frac{G_N m}{\omega d^3}  \approx \SI{6e-8}{\hertz} \times \left( \frac{m}{\SI{1}{\nano\gram}}\right) \left( \frac{\SI{1}{\hertz}}{\omega} \right) \left( \frac{\SI{1}{\micro\meter}}{d} \right)^3.
\ee
Here $x_{ZPF} = \sqrt{\hbar/2 m \omega}$ is the ground-state uncertainty in the harmonic oscillator position. Note that the ratio $m/d^3$ can be no larger than the material density of the resonators.

The Hamiltonian \eqref{2osc-H} manifestly generates entanglement between the oscillators. As a simple example, consider preparing one oscillator in its ground state and the other in its first excited Fock state, that is $\ket{\psi} = \ket{10}$. Treating the gravitational interaction as a perturbation, this state will evolve to
\be
\label{10example}
\ket{10} \to \ket{10} - i \lambda_g \Delta t \ket{01} + \Oi(\lambda_g^2)
\ee
in a time $\Delta t$, up to an overall normalization. The state \eqref{10example} is entangled, in the sense that it cannot be reduced to a product state of the two oscillators. The amount of entanglement is small, since it is proportional to the product $\lambda_g \Delta t$, but it could in principle be measured by a DLCZ-style scheme \cite{duan2001long}. For example, if these oscillators are mirrors in an optical cavity, we can apply a red-detuned optical pulse, destroying the joint oscillator phonon state and mapping it onto a pair of optical photons, whose interference visibility can be measured using standard optics techniques. This technique has been successfully used to demonstrate entanglement in a pair of oscillators in the evenly-weighted state $\ket{10} + \ket{01}$ generated using photon-phonon couplings\cite{riedinger2016non}.

A more dramatic example would be to prepare the source oscillator not in a low-lying Fock state like $\ket{1}$ but a highly non-classical state like a ``cat code'' $\ket{\alpha} + \ket{-\alpha}$.\cite{gottesman2001encoding} Here $\ket{\alpha}$ is a coherent state, which has average spatial displacement $\braket{ x } = \text{Re}(\alpha)$ from the oscillator equilibrium. Consider preparing the first oscillator in this state and the second oscillator in its ground state $\ket{0}$. We can easily find the exact time-evolution of the joint state without resorting to perturbation theory. Defining as usual the normal modes
\be
x_{\pm} = \frac{x_1 \pm x_2}{\sqrt{2}}
\ee
the Hamiltonian can be written
\be
\label{H-2osc-rwa}
H = \omega_+ a_+^{\dagger} a_+ + \omega_- a_-^{\dagger} a_-, \ \ \ \omega_{\pm} = \omega \pm \lambda_g.
\ee
Using this transformation, it is straightforward to show that a coherent state times the ground state evolves as
\begin{align}
\begin{split}
\ket{\alpha}_1 \ket{0}_2 & \approx \ket{\alpha/\sqrt{2}}_+  \ket{-\alpha/\sqrt{2}}_- \\
&  \to \ket{e^{-i \omega_+ t} \alpha/\sqrt{2}}_+ \ket{-e^{-i \omega_- t} \alpha/\sqrt{2}}_- \\
& \approx \ket{\alpha_t \cos \lambda_g t}_1 \ket{-i \alpha_t \sin \lambda_g t}_2.
\end{split}
\end{align}
Here we defined $\alpha_t = \alpha  e^{-i \omega t}$, and approximated $\omega_{\pm} \approx \omega$ in the widths of the ground states of the $\pm$ modes; this creates errors in the final answers only at second order in $\lambda_g$. Extending this calculation by linearity, we see that the initial cat-vacuum state
\be
\ket{\psi} = \frac{\ket{\alpha} + \ket{-\alpha}}{\sqrt{2}} \otimes \ket{0}
\ee
will evolve to
\be
\label{2-osc-state-final}
\ket{\psi(t)} = \left( \ket{\alpha_t \cos \lambda_g t}\ket{-i \alpha_t \sin \lambda_g t} + \ket{-\alpha_t \cos \lambda_g t} \ket{i \alpha_t \sin \lambda_g t} \right)/\sqrt{2}.
\ee
This state is a highly non-classical entangled state; for large $\text{Re}(\alpha)$ the two branches are essentially orthogonal. This also plainly demonstrates the transfer of quantum information from one oscillator into the other: for example, at $t = \pi/2\lambda_g$, the initial state of the first oscillator has been completely swapped (up to some phases) into the second oscillator.

The above examples show how the Newtonian force in the EFT picture of gravity acts as a ``quantum'' interaction. What this means precisely is that the interaction is capable of transmitting quantum information: information about a superposition of one system can be mapped into another system, and so forth. This is in stark contrast to known models in which a classical gravitational field somehow couples to quantum matter, as we will see in section \ref{classical-gravity}. More technically, this is the statement that the Newtonian interaction $V = Gm^2/|r_1-r_2|$ is an \emph{operator} acting on the joint Hilbert space of two masses. This is a simple consequence of two assumptions: that we quantize matter, and that the gravitational field is treated as a quantum field sourced by matter in accordance with EinsteinÕs field equations and the principle of equivalence. Because of the latter assumption, the gravitational field is constrained to follow the matter field through the gravitational Gauss law, and so it must engage in any superpositions the matter field happens to be involved in.\footnote{One could in principle ask if there are any consistent theories in which the Newton interaction can generate entanglement in this way, but in which there are no quantized graviton fluctuations; under very simple assumptions, the answer appears to be no, see \cite{Belenchia:2018szb} and references therein.}

\begin{figure}[t!]
\begin{center}
\includegraphics[scale=1.2]{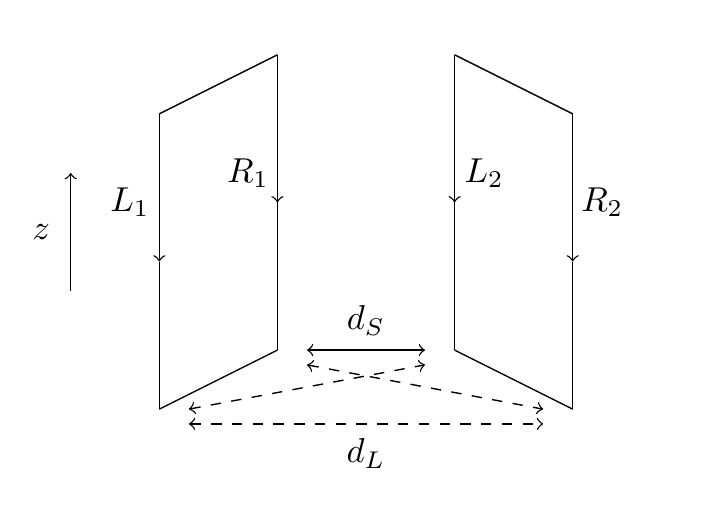}
\end{center}
\caption{Schematic of the proposal \cite{bose2017spin,marletto2017entanglement} for testing gravitationally-generated entanglement. A pair of matter-wave interferometers are placed next to each other. Here we have picked a particular geometry in which the matter waves will be free falling, and such that the $R_1 - L_2$ arms have a short separation $d_S$ while the other three combinations have a long separation $d_L > d_S$.}
\label{bose-figure}
\end{figure}

Having explained the basic idea of gravitational entanglement generation using resonator systems, we now remark briefly on a recent, concrete proposal to study the same phenomenon in a matter-wave system \cite{bose2017spin,marletto2017entanglement}. Consider a pair of matter-wave interferometers placed next to each other as in figure \ref{bose-figure}. The matter waves consist of massive objects which have a spin-1/2 degree of freedom which can be used to both control and read out the spatial state of the masses; in the proposal \cite{bose2017spin} the masses are nitrogen-vacancy diamonds (with the NV site providing the spin) and the ``beam splitting'' action is accomplished by engineered magnetic fields coupling to the spin \cite{pino2018chip}. We denote the beam arms $\ket{L} = \ket{\uparrow}, \ket{R} = \ket{\downarrow}$ in the notation of section \ref{simplemodels}. Neglecting all interactions except for the Newtonian interaction between the two matter waves, the total time evolution produces a state of the form
\be
\label{bose-result}
\ket{LL} \to e^{i \phi_{LL}} \ket{LL} + e^{i \phi_{LR}} \ket{LR} + e^{i \phi_{RL}} \ket{RL} + e^{i \phi_{RR}} \ket{RR},
\ee
with the phases $\phi_{ij} = G_N m^2\Delta t/\hbar d_{ij}$, $d_{ij}$ the spatial distances between the various pairs of beam arms. For simplicity consider a geometry for the interferometers such that $\phi_{LL} = \phi_{RR} = \phi_{LR} \neq \phi_{RL}$ (for example, the geometry in figure \ref{bose-figure}), so that we may write this as
\be
\ket{LL} \to \ket{LL} + \ket{LR} + e^{i \Delta \phi} \ket{RL} + \ket{RR},
\ee
up to an overall phase, with the differential phase of order
\be
\Delta \phi = \frac{G_N m^2 \Delta x \Delta t}{\hbar d^2} \approx 60 \times \left( \frac{m}{\SI{1}{\nano\gram}} \right)^2 \left( \frac{\Delta x}{\SI{1}{\micro\meter}}\right) \left( \frac{\Delta t}{\SI{1}{\second}} \right) \left( \frac{\SI{1}{\milli\meter}}{d} \right)^2.
\ee
Here $d$ is the separation between the two interferometers, $\Delta x$ the separation between the beam arms in a single interferometer, $\Delta t$ the time of flight, and we have taken the approximation $\Delta x/d \ll 1$ (in terms of the geometry in figure \ref{bose-figure}, we are approximating $d_S \approx d, d_L \approx d + \Delta x$). This is manifestly an entangled state of the two masses: it is not a product state.\footnote{Note that this is not true if $\phi_{LR} = \phi_{RL} + 2\pi n$ for any $n$. Thanks to Michael Graesser and Daniel Oi for discussions on the particulars of the geometry needed for the experiment.} The entanglement can be verified by measuring the spin states, for example by looking for violations of a Bell inequality: if the entanglement witness
\be
W = | \braket{ \sigma_x^1 \sigma_z^2 - \sigma_y^1 \sigma_z^2} | \leq 1
\ee
then the state is provably entangled.

\subsection{Tests of graviton-induced decoherence channels}
\label{gravitons}

One of the most striking predictions about perturbatively quantized general relativity is the existence of low-energy fluctuations in the field: gravitons. There are strong arguments that it will be impossible to ever directly detect a single graviton, even in principle--it would generally require a detector so large and massive that a black hole would form \cite{dyson2013graviton,rothman2006can,boughn2006aspects}. However, there is an alternative: perhaps one could infer the existence of the graviton, or more precisely the ability of systems to spontaneously emit gravitons, by observing decoherence consistent with information lost into graviton states \cite{anastopoulos2013master,blencowe2013effective,oniga2016quantum}.

Unfortunately, as we review in this section, the rates of graviton-induced decoherence are extraordinarily low compared to more pedestrian channels like electromagnetic radiative damping, at least for small masses. Our primary purpose here is to explain the nature of these effects ``in principle'', and in particular to demonstrate that standard, quantized gravitons can lead to decoherence in precise analogy with photons. This decoherence must be distinguished from the ad hoc ``gravitational decoherence'' mechanisms discussed in section \ref{alt-grav-models}, which generally speaking have enormously higher rates than that coming from gravitons.

To get a feel for the relative size of photonic versus gravitonic effects, we can estimate the rates of spontaneous emission of either a photon or a graviton. The rate for spontaneous emission of photons should depend on the system's electric dipole made into a scalar $\mb{d}^2$, the speed of light $c$, Planck's constant $\hbar$, and the system's frequency of oscillation $\omega$; forming a rate from these quantities, on dimensional grounds, gives the estimate 
\be
\Gamma_{EM} \sim \frac{d^2 \omega^3}{\hbar c^3}.
\ee
For spontaneous emission of a graviton, since we are dealing with spin-2 emission instead of spin-1, we should have a dependence on the quadrupole moment $Q^2$, and furthermore we should expect a factor of $G_N$ from the graviton coupling to matter. Again dimensional analysis gives
\be
\Gamma_{GR} \sim \frac{G_N Q^2 \omega^5}{\hbar c^5}.
\ee
To get a concrete estimate, we can approximate the dipole as consisting of $N$ elementary dipoles, $d^2 = N^2 \alpha \hbar c L^2$, with $L$ the linear dimension of the object and $\alpha = e^2/\hbar c = 1/137$ the fine structure constant. Similarly we approximate the mass quadrupole $Q = m L^2$. Then we have
\be
\label{em-gr-power}
\frac{\Gamma_{EM}}{\Gamma_{GR}} \sim \frac{N^2 \alpha \hbar c^3}{G_N m^2 L^2 \omega^2}.
\ee
To get a feel for this, consider an object with $L \sim \SI{1} {\micro\meter}, \omega \sim \SI{1}{\mega\hertz}, m \sim \SI{1}{\nano\gram}$; this gives $\Gamma_{EM}/\Gamma_{GR} \sim N^2 \times 10^{23}$. In other words, even for a fairly aggressive choice of experimental parameters, the electromagnetic radiation absolutely swamps any graviton-based effects. However, it should be noted that for very large masses the gravitational rate can actually start to dominate \cite{reynaud2001gravitational}. Beyond this order-of-magnitude estimate, we refer the reader to detailed calculations of spontaneous emission in a neutron quantum bouncer \cite{pignol2007spontaneous}, simple harmonic oscillator \cite{oniga2016quantum}, and especially the wonderful review of the $3d \to 1s$ transition in hydrogen presented in \cite{rothman2006can,boughn2006aspects}.

Another possibility would be to look for thermalization due to an ambient graviton background. Indeed, little is known about the precise nature of the ambient graviton background, and observing decoherence due to this background would be remarkable, to put it mildly. A simple estimate suggests that, if the early universe went through a period of inflation, a background of thermalized gravitons at $T \lesssim \SI{1}{\kelvin}$ should be universally present, if nothing else \cite{kolb2018early}. Stars, in particular the sun, also emit gravitons at reasonable rates; a simple estimate leads to a flux of about $10^{-4}$ gravitons per square centimeter per second on the Earth \cite{weinberg1965infrared,dyson2013graviton}. To estimate the thermalization rate, we can again use dimensional analysis: the rate $\Gamma^{dec} \sim c n_{grav} \sigma$, where $n_{grav}$ is the number density of the graviton background, and $\sigma$ is an effective graviton absorption cross-section. The cross section can be conservatively estimated by $\sigma \sim 4\pi^2 L_p^2 \sim \SI{e-65}{\centi\meter^2}$ with $L_p$ the Planck length \cite{dyson2013graviton}, so for example solar gravitons lead to a decoherence rate $\Gamma^{dec} \sim \SI{e-69}{\hertz}$, which is utterly negligible.

One might also consider decoherence caused by the bremsstrahlung of gravitons during the acceleration of a massive object. For example, imagine using a matter-wave beam of particles of mass $m$ and sending these through a beam-splitter. We can think of the beam-splitter as the unitary which implements a simple $90$-degree rotation of the incoming momentum vector:
\be
\ket{ p \hat{x} } \to \ket{ p \hat{x} } + \ket{ p \hat{y} }
\ee
where the kets are momentum eigenstates with momenta $p$ aligned along either the $x$ or $y$ axis; see figure \ref{cartoon-mw}. The process $p \hat{x} \to p \hat{y}$ involves acceleration of the beam, and thus will generate bremsstrahlung in both gravitons and, if the beam is electrically charged, photons. Thus in actuality, the  beamsplitter acts as
\be
\ket{ p \hat{x} , 0 } \to  \ket{ p \hat{x}, 0 } + \sum_{\gamma,h} c_{\gamma,h} \ket{ p \hat{y} , \gamma,h } 
\ee
where the $0$ indicates the vacuum state of the radiation field, $\gamma,h$ represent some possible photon and graviton radiation, and the $c_{\gamma,h}$ are the amplitudes for each of these states. 

The radiation spectrum has a divergent infrared tail: an arbitrarily small amount of energy can be radiated out into an arbitrarily large number of very low-energy bosons\cite{jackson1975electrodynamics,bloch1937note,weinberg1965infrared}. These in turn cannot be measured by a finite-energy detector, and should be traced. This divergence is regulated by the temporal duration $\tau$ of the experiment, because a boson with wavelength $\lambda \gtrsim c \tau$ cannot be produced. One can perform the trace to all orders in the number of photons and gravitons \cite{carney2017infrared}, and one finds that the off-diagonal component of the density matrix decays like
\be
\rho_{LR}(\tau) \sim \rho_{LR}(0) \exp \left\{ - \left( \frac{p}{P_{p}} \right)^2 \ln \left( \frac{m c^2 \tau}{\hbar} \right) \right\}.
\ee
Here $P_{p} = M_{p} c \approx \SI{6}{\kilo\gram \meter \per\second}$ is the Planck momentum, and we have ignored the photon contributions--this is the pure graviton decoherence rate. For a typical matter-wave experiment we have $m \sim \SI{e5}{\amu}, \tau \sim \SI{1}{\second}, v \sim \SI{10}{\meter \per \second}$, and so this exponential is astronomically close to being equal to unity. In other words, the decoherence is entirely negligible. To get a non-negligible contribution, one basically needs $p/P_{pl} \sim \Oi(1)$, i.e. momenta of order $\SI{1} {\kilo\gram \meter \per \second}$, which seems rather difficult to achieve. One can compare this with the electromagnetic version, in which we replace $p/P_{pl} \to N^2 \alpha$ with $\alpha$ the fine structure constant and $N e$ is the charge of the beam particles; in this case, it is easy to imagine an experiment with non-negligible decoherence from photon bremsstrahlung.

Clearly, observing decoherence caused by gravitons will be quite challenging. Still, there seems to be room for new ideas: although detecting a single graviton can be fairly conclusively deemed impossible on simple grounds, here we have only sketched some very rudimentary schemes for decoherence-based detection. One obvious avenue may be to engineer a system with a particularly large mass quadrupole and minimal electromagnetic couplings. Another could be to exploit a Dicke-superradiant state of some system to enhance the spontaneous graviton emission rate \cite{quinones2017quantum}. We look forward to the development of new ideas in this vein, but for the time being, now turn our attention to gravitational models beyond the effective quantum field theory picture of general relativity.

\section{Gravity as a fundamentally classical interaction}
\label{classical-gravity}

In the previous section, we studied the low-energy predictions for quantum general relativity, considered as an effective quantum field theory. Broadly speaking, this would be the standard picture in which gravity at low energies behaves quantum mechanically. Standard quantum field theory is taken as axiomatic, and classical gravity emerges as a limit of quantum gravity. In the classical limit, massive bodies interact through a classical Newtonian force, and the metric fluctuations are classical gravitational waves.

However, to date we have no direct evidence that gravity behaves quantum-mechanically, and are thus obliged to consider alternatives. In this section, we consider what it might mean for gravity to be ``fundamentally classical''. This could potentially mean a number of things, and here we will study two definite proposals. We begin with a model which we will refer to as ``traditional semiclassical gravity'', emphasizing its theoretical inconsistencies. We then move on to a model of the gravitational interaction as a classical information channel, showing how this model can reproduce the basic intuition of semiclassical gravity within a theoretically consistent framework. In both cases, a key prediction is that the gravitational interaction is incapable of transmitting any quantum information between masses, and in particular cannot generate entanglement, in stark contrast to the quantum EFT picture of section \ref{EFTsection}.

Suppose we want to treat gravity classically and couple it to the quantum state of matter. The simplest way we could try to implement this is by sourcing the Einstein equations with the expectation value of the matter stress energy tensor:
\be
\label{scg-1}
G_{\mu\nu} = \frac{8 \pi G_N}{c^4} \braket{ T_{\mu\nu}}.
\ee
Here, $\braket{T_{\mu\nu}} = \braket{ \psi | T_{\mu\nu} | \psi}$ is the expectation value of the quantum-mechanical stress tensor for the matter, and this equation determines the dynamics of the spacetime metric. The theoretical troubles begin when one attempts to self-consistently close this system with a Schr\"{o}dinger equation for the matter
\be
\label{scg-2}
i \partial_t \ket{\psi} = \left( H_{mat} + H_{grav} \right) \ket{\psi},
\ee
where $H_{grav}$ is the gravitational potential encoded in the semiclassical Einstein equation \eqref{scg-1}. These equations are sometimes referred to as ``semiclassical gravity'', and sometimes as the ``Schr\"{o}dinger-Newton'' model, especially in the case that $H_{mat}$ describes non-relativistic particles \cite{ROSENFELD1963353,moller1962theories,Kibble:1978vm,Kibble:1979jn,Adler:2006yj,Salzman:2006rk,Carlip:2008zf,anahusn,Bahrami:2014gwa}. 

The term ``semiclassical gravity'' may cause confusion, because it can refer to two rather different ideas. Typically one has a perfectly ordinary quantum system, and a ``semiclassical'' treatment is one in which we have an expansion in powers of $\hbar$, i.e. an expansion in quantum fluctuations, around a classical limit. In this sense, the semiclassical equations \eqref{scg-1} and \eqref{scg-2} are a perfectly valid limit of the full EFT equations in the limit that the quantum stress tensor has only small fluctuations, but this is precisely the opposite of the situation considered in this paper, in which the matter is prepared in a highly quantum state. 

Conversely, here and throughout this section, we will take ``semiclassical gravity'' (and its non-relativistic limit, the Schr\"{o}dinger-Newton model) to mean something radically different: we simply take equations \eqref{scg-1}, \eqref{scg-2} as fundamentally true. In this sense we are studying a departure from the EFT picture. Although this may seem like a plausible candidate for a theory of classical gravity coupled to quantum matter, there are severe and fundamental problems with this idea. In particular, since the gravitational potential \eqref{scg-1} depends on the matter state $\ket{\psi}$, and this potential in turns determines the evolution of the matter state in \eqref{scg-2}, this amounts to a non-linear modification of the usual quantum time evolution. There are then two ways to view the semiclassical model: as a fundamental modification to standard quantum theory, or as a kind of flawed limit of some more reasonable theory of ``classical'' gravity coupled to quantum matter. In section \ref{scg}, we take the former view; in this section, we instead show how to embed the semiclassical equations into a consistent quantum model, in which the non-linearity arises in a controlled fashion.

Before giving the full construction, we can study the basic predictions of such a semiclassical model, ignoring the theoretical inconsistencies for a moment. In section \ref{scg} we give a more extensive list of experimental proposals for testing the Schr\"{o}dinger-Newton equation; here we focus on a simple two-body entanglement experiment to contrast the results with those for the EFT picture of section \ref{EFTsection}. 

The point we would like to emphasize is that one consequence of equations \eqref{scg-1} and \eqref{scg-2} is that the gravitational field cannot transmit quantum information between matter. In particular, the gravitational field cannot cause entanglement. To see this, we can look at the non-relativistic limit of these equations. The Einstein equations reduce to the gravitational Poisson equation for the Newtonian potential $\Phi$,
\be
\label{scg-nr1}
\nabla^2 \Phi = 4 \pi G_N \braket{M},
\ee
where $M = M(\mb{x})$ is the mass density operator of the matter (we use $M$ instead of $\rho$ to avoid confusion with the quantum-mechanical density matrix). In the non-relativistic limit with $N$ particles, this operator can be written
\be
\label{massdensityop}
\hat{M}(\mb{x}) = \sum_i m_i \delta(\mb{x} - \hat{\mb{x}}_i)
\ee
where for clarity we use a hat to denote an operator; the sum over $i=1,\ldots,N$ is a sum over the particles.\footnote{To derive this expression, one can take a scalar field $\phi$ for the matter content, and study the non-relativistic limit of $T_{00}$ in a single-particle state of $\phi$. The operator $T_{00}(x) \sim (\partial \phi(x))^2$ has an ultraviolet divergence because it has two operator insertions at a single point $x$. This divergence must be regulated and renormalized, and so the parameters $m_i$ here are really the renormalized masses of the particles.} The gravitational interaction with matter in this limit is given by $H_{grav} = \int d^3\mb{x} M(\mb{x}) \Phi(\mb{x})$, and so the Schr\"{o}dinger equation reduces to
\be
\label{scg-nr2}
i \partial_t \ket{\psi} = \left( H_{mat}  + \int d^3\mb{x} \ M(\mb{x}) \Phi(\mb{x}) \right) \ket{\psi},
\ee
where the potential satisfies \eqref{scg-nr1}. 

Consider a pair of massive particles, for example the matter-wave beams in the twin-interferometer experiment of figure \ref{bose-figure}. In section \ref{eft-newton} we explained how the particles in the two interferometers will become entangled with each other through the quantum Newton interaction, leading to the final state \eqref{bose-result}. In the semiclassical model presented here, the dynamics are quite different: each of the four interferometer arms sees an effective classical potential sourced by all three of the other arms, through the expectation value $\braket{\rho(\mb{x})}$. This can be computed as
\be
\braket{ \rho(\mb{x}) } = \frac{m}{2} \left[ \delta(\mb{x}-\mb{x}_{1,L}) + \delta(\mb{x}-\mb{x}_{1,R}) + \delta(\mb{x}-\mb{x}_{2,L}) + \delta(\mb{x}-\mb{x}_{2,R}) \right],
\ee 
if we approximate the particles in the interferometer arms as position eigenstates $\ket{\mb{x}_{i,L/R}}$, with $i=1,2$ labeling the two interferometers. No entanglement is produced, and a simple calculation shows that we obtain instead the final state
\be
\ket{LL} \to \left( \ket{L} + e^{i \Delta \phi} \ket{R} \right)_1 \otimes \left( \ket{L} + e^{-i \Delta \phi} \ket{R} \right)_2,
\ee
where the relative phase $\Delta \phi$ is the same as given in \eqref{bose-result}, and we have again ignored any changes in the kinetic energies of the masses. This is a product state and, in particular, would never lead to a violation of a Bell inequality in a joint measurement of the two particles. Nothing here is special to the interferometer case; we could just as well try to look for entanglement generated between a pair of resonators. While the EFT model of gravity predicts entanglement in both cases, the classical model does not.

\begin{figure}[t]
\begin{center}
\includegraphics[scale=0.9]{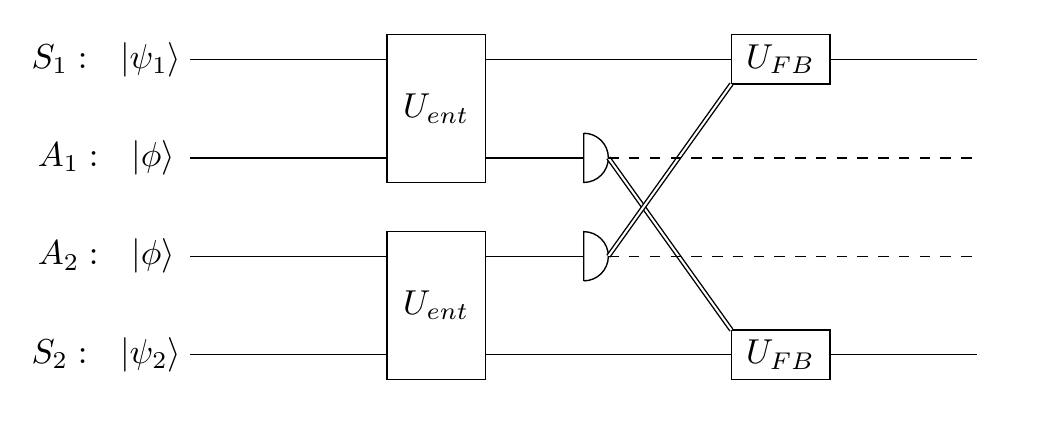}
\end{center}
\caption{Circuit diagram for the classical channel model of gravitational interactions, in a two-body example. The lines $S_1,S_2$ correspond to the two massive bodies and the lines $A_1,A_2$ are the ancillae used to measure the positions of the system. The unitaries $U_{ent}$ entangle the matter system with the ancillae; the caps represent measurement. This is used to estimate $\braket{T_{\mu\nu}}$, and this classical information is then fed back onto the matter system with $U_{FB}$. No quantum information flows between the matter systems.}
\label{kmt-fig}
\end{figure}

Although this model has some utility as a heuristic picture of classical gravity coupled to quantum matter, one would prefer a more theoretically sound alternative. All of the difficulties arise from non-linearity in the Schr\"{o}dinger equation, which was imposed as a fundamental property. Here we present an approach which evades these problems by positing a mechanism for the non-linearity: we imagine some ancilla system which continuously monitors the matter stress-energy, using its measurements to estimate $\braket{T_{\mu\nu}}$. The ancilla then acts to classically feed this information back onto the matter, such that a semiclassical gravitational interaction arises. In this sense the gravitational interaction becomes a classical information channel, which is incapable of transmitting quantum information. The price paid is an irreducible amount of noise arising from the measurements, but since the entire model including the ancillae is ultimately unitary and quantum, we evade the more serious theoretical consistency issues \cite{ralph2009quantum,Kafri:2014zsa,Kafri:2015iha,tilloy2016sourcing,Altamirano:2016knc,Tilloy:2018tjp}. The idea of using noise terms to self-consistently complete semiclassical gravity goes back to Diosi \cite{diosi1986universal}; here we are providing a manifestly unitary, microscopic model of the origin of the noise.

Time evolution in this measurement-and-feedback model is most easily described in terms of timesteps $\Delta t$; we will take the continuum limit at the end. For concreteness, we will work in a non-relativistic limit, and assume each particle in the model has the same mass $m$. Each timestep has two distinct processes; see figure \ref{kmt-fig} for a circuit diagram in a simple two-body example. First, the matter system has its mass density weakly (i.e. non-projectively) measured, for example by entangling the matter with some ancilla system and projectively measuring the ancilla. According to the usual von Neumann postulate, this causes a change in the matter state $\ket{\psi} \mapsto \ket{\psi} + \Delta \ket{\psi}$, which becomes conditioned on the outcome of the measurement:
\be
\label{conditionedstate}
\Delta \ket{\psi} = \left[ \int d^3\mb{x} \xi(\mb{x}) \sqrt{\gamma} \Delta W(\mb{x}) - \frac{1}{2} \int d^3\mb{x} d^3\mb{y} \xi(\mb{x}) \xi(\mb{y}) \gamma \Delta t \right] \ket{\psi}.
\ee
Here, the parameter $\gamma$ controls the strength of the measurement; in line with the usual continuous weak measurement paradigm \cite{Caves:1987zz,jacobs2006straightforward,ZHANG20171,caves2}, we take $\gamma \Delta t \ll 1$, and have performed a Taylor expansion in this quantity. The measure $\Delta W(\mb{x})$ is a stochastic random variable, which accounts for the measurement outcomes; in order to satisfy the usual stochastic calculus it should be normalized to $\text{var}\ \Delta W = \Delta t$. Finally, the operator $\xi(\mb{x})$ is
\be
\xi(\mb{x}) = \frac{M(\mb{x}) - \braket{M(\mb{x})}}{M_{p}},
\ee
where $M(\mb{x})$ is the mass density operator \eqref{massdensityop}, so $\xi(\mb{x})$ represents the deviation from the average mass density, measured in units of the Planck mass $M_{p}$. Note that the operator $M(\mb{x})$, which involves pointlike masses, can generally lead to divergences; these should be regulated by introducing some fundamental ultraviolet regulator like a lattice spacing $R_0$.

The evolution \eqref{conditionedstate} is non-linear in that it depends on expectation values taken in $\ket{\psi}$. However, this non-linearity is now understood as arising from a series of measurements done on some ancilla systems, and in this sense the complete matter plus ancilla system is described within standard quantum mechanics, so there is no fundamental modification of the basic quantum formalism.

The second part of the timestep is to use this measurement data to feed back a classical gravitational force on the system. The measurement outcome is used to estimate the mass density $\braket{M(\mb{x})}$, which in turn is used to find the Newtonian potential $\Phi$ satisfying the Poisson equation $\nabla^2 \Phi = 4 \pi G_N \braket{M(\mb{x})}$. We then simply act on the system with the unitary
\be
U_{feedback} = \exp \left\{ -i \Delta t \int d^3\mb{x} M(\mb{x}) \Phi(\mb{x}) \right\}.
\ee
Putting this together with the evolution from the measurement \eqref{conditionedstate} and taking the continuum limit, we have a total change in the matter state given by
\begin{align}
\label{kmt-final}
\begin{split}
d \ket{\psi} = \Big\{ & -i H_{mat} d t - i \int d^3\mb{x} M(\mb{x}) \Phi(\mb{x}) d t \\
&  + \left[ \int d^3\mb{x} \xi(\mb{x}) \sqrt{\gamma} dW(\mb{x}) - \frac{1}{2} \int d^3\mb{x} d^3\mb{y} \xi(\mb{x}) \xi(\mb{y}) \gamma d t \right] \Big\} \ket{\psi},
\end{split}
\end{align}
where $H_{mat}$ is the non-gravitational Hamiltonian. The terms in the first line generate precisely the semiclassical gravitational evolution \eqref{scg-2}, while the terms in the second line represent an irreducible source of noise from the continuous measurement of the matter system. In particular, averaging over the measurement histories $dW$ will lead to dephasing, so this model essentially includes an extra decoherence channel from these measurements.

The time-evolution \eqref{kmt-final} has some important features. It reproduces the semiclassical gravitational interaction of the Schr\"{o}dinger-Newton model, without any basic theoretical difficulties; in particular, there is no violation of the no-signalling condition. Like the SN equation, this evolution generates no entanglement between the masses, so either model can be falsified by a two-body gravitational entanglement measurement. The price we pay for self-consistency is the introduction of the noise $dW$; averaging over this measurement noise will lead to dephasing of the matter. Among other things, this will result in anomalous heating in any object. Observations of long-lived Bose-Einstein condensates, for example, have been used to place numerical bounds on the parameter $R_0$ in this model\cite{Kafri:2015iha}. Similar bounds can be placed using a long-lived resonator or matter-wave beam. 

There is significant structural freedom in models of this type: one can choose different ancillae, measurements, etc. For example, one could consider minimizing the anomalous heating rate just mentioned \cite{Kafri:2015iha}, or alternatively minimizing decoherence rates \cite{tilloy2017principle}. Certain highly-simplified versions of this proposal can already be subjected to experimental test \cite{altamirano2018gravity}, but a detailed, relativistic, field theory-based model is still lacking. Upgrading this model to a fully relativistic setting will be challenging: how to treat the measurements in a manner consistent with coordinate invariance and how to incorporate the gravitational effects of the ancillae themselves are topics for future work. Nevertheless, as a concrete model for gravity as a fundamentally classical interaction in the non-relativistic regime, this seems to be the most promising candidate available.

Before moving on from the notion of classical gravity, we remark on one further, highly interesting possibility. It has been suggested that classical or semiclassical gravity could be ``emergent'', in the sense that it is a long-wavelength limit of some underlying, unknown quantum degrees of freedom, not necessarily gravitons \cite{Jacobson:1995ab,Verlinde:2010hp,Hossenfelder:2010ih,padmanabhan2015emergent,Verlinde:2016toy,Hossenfelder:2017eoh}. This should be viewed roughly in analogy with the hydrodynamic limit of fluid mechanics, in which the underlying quantum atoms lead to semiclassical hydrodynamics at sufficiently long length scales. A variety of such models have been proposed; and their predictions for experiments like those we are studying here are unclear, although there are simple models demonstrating that, entropic forces cannot entangle particles \cite{wang2016quantum} and there has been some debate \cite{Kobakhidze:2010mn,Chaichian:2011xc} about whether or not these models are consistent with experiments involving gravitationally-bound cold neutron states \cite{nesvizhevsky2002quantum}. We view the types of experiments discussed in this paper as a major opportunity to learn more about such emergent models. We leave this idea to future work.

\section{Low-energy gravity models violating quantum mechanics}
\label{alt-grav-models}

So far, we have studied gravity as a perturbative quantum field theory treated much the same as electrodynamics, and contrasted this with some models of gravity as a purely classical interaction. We argued that one possibly consistent way to view gravity as classical is as a kind of limit of a perfectly unitary quantum model, in which some ancillary system is used to enact semiclassical gravitational interactions. 

In this section, we turn instead to some models built out of the notion that gravity and quantum mechanics are fundamentally incompatible in some way. This is a point of view that goes back to Einstein \cite{einstein50}, and which has been steadily refined since the late 1950's. To date, gravitational effects are negligible at the microscopic scales where all entanglement and Bell inequality experiments have been done; even experiments in which ``macroscopic state superpositions" of flux \cite{chiorescu2003coherent} are created\footnote{Although there is considerable debate \cite{korsbakken2007measurement,korsbakken2010size,volkoff2014measurement,leggett2007probing,knee2016strict,leggett2016note} about how macroscopic these superpositions are.} do not involve any significant mass displacements, and so no gravitational effects are involved there either. Thus, for various reasons to be described below, one can entertain the idea that quantum mechanics may break down at large scales, and that gravitation may be involved. 

There are various such theories: some involve semiclassical approaches in which gravity is sourced by expectation values of the stress-energy tensor as described above, and others invoke other ways of violating the superposition principle. All of them go outside the EFT framework discussed above. Instead one is now looking for violations of quantum mechanics, taking place when one tries to superpose or entangle states which are macroscopically different in their gravitational properties.

The key challenges are then to find consistent alternative theories which mimic quantum theory at small scales, but violate it at large scales, because of gravitational effects; and to find experimentally accessible phenomena in which the difference between these theories and standard quantum mechanics will show up. For theories of this kind to be convincing, they need to be predictive, and this means they need to involve a clear physical mechanism. In what follows we focus on theories which may lead to experimental tests.

\subsection{Early lessons from non-linear theories}
 \label{scg}

The first attempts to grapple with the challenges noted above were in early remarks by Feynman \cite{cecile2011role,feynman1971lectures}, and in an analysis of uncertainty relations involving masses coupled to gravity by Karolyhazy \cite{karolyhazy1966gravitation}. However, no attempt was made in this early work to provide any kind of alternative theory. The first serious attempt at such a theory was made by Kibble et al. in \cite{Kibble:1978vm,Kibble:1979jn}. Quite apart from the details, three key points were made by Kibble et al., which have just as much force today. These were:

(i) Any semiclassical theory of gravity (in which the matter fields are all quantized, but the metric is taken to be classical in a sense to be specified) will differ in its predictions from a fully quantized theory in which both the matter fields and the metric are quantized together.

In the literature, a semiclassical theory is, as before, typically taken to mean one for which the relation between the Einstein tensor and the stress-energy tensor is given by
\be
\label{scg-eq1}
G_{\mu\nu} = \frac{8 \pi G_N}{c^4} \braket{ T_{\mu\nu}}.
\ee
in which $\braket{ T_{\mu\nu}}$ is the expectation value of the quantum-mechanical $ T_{\mu\nu}$. A simple example of the predictions for such a model were given in the previous section; let us put this in more general terms. Consider a two-slit experiment, in which the two paths for a mass $M$ are significantly separated in space. In a theory with a quantized metric, such a superposition would take the form
\begin{equation}
\ket{\Psi} = a_L \ket{\Phi_L; g^{(L)}_{\mu \nu}} + a_R \ket{\Phi_R; g^{(R)}_{\mu \nu}},
 \label{GRsup}
\end{equation}
with $L,R$ denoting the two paths. In this superposition, those gravitational degrees of freedom that are tied to matter in the gravitational part $\ket{g_{\mu \nu}}$ of the state vector are then completely entangled with the matter state $\ket{\Phi}$. If $M$ is small enough to treat these metrics as perturbations around flat spacetime, then this description is just what was given in the EFT picture above.

Suppose we try to measure which path the mass $M$ has followed using, for example, another mass $m$ to measure the metric perturbation associated with the mass $M$. It is immediately obvious that in a fully quantized theory, the secondary ``apparatus" mass $m$ will be deflected differently depending on which path the mass $M$ follows: we will end up with a superposition
\begin{equation}
 \label{GRsup2}
\ket{\Psi_f} = a_L \ket{\Phi_L; g_{(L)} ; \chi_{(L)}} + a_R \ket{\Phi_R; g_{(R)} ;\chi_{(R)}}
\end{equation}
where the apparatus mass states $\ket{\chi_{(L)}}$, $\ket{\chi_{(R)}}$ have the mass $m$ deflected in different directions by the fields $g_{(L)}$, $g_{(R)}$. A standard projective measurement done jointly on the two masses would then see the apparatus deflected one way or the other in correlation with the source, with probabilities $|a_L|^2$ and $|a_R|^2$ respectively.

On the other hand in a semiclassical theory where \eqref{scg-eq1} is satisfied, we get a quite different result. The expectation value of the source mass's stress tensor is a classical sum of two locations. For example, suppose that $a_L = a_R = 1/\sqrt{2}$ in \eqref{GRsup} and the apparatus is placed directly between the two source paths $L,R$. Then the apparatus mass will be pulled equally toward each path, and will remain undeflected. In particular, absolutely no entanglement is generated between the source and apparatus, as discussed above.

Other writers have commented on this point of Kibble's \cite{page1981indirect,unruh86}, and it was even claimed \cite{page1981indirect} that measurements have already decided in favour of the superposition \eqref{GRsup}. This claim is controversial for several reasons \cite{hawkins1982indirect,ballentine1982comment,page1982page}; from our point of view the chief problem is that no sources of environmental decoherence sensitive to the path of the mass $M$ are considered in any of the above, and such decoherence (which can come from many sources, including photons, phonons, and gas atoms) is likely to be large in any experiment where $M$ is large enough to have appreciable gravitational effects.

(ii) Any semiclassical theory of gravity must necessarily involve non-linear time evolution in the matter fields, and hence break the superposition principle. This means, for example, that in the non-relativistic limit we will be dealing with a generalized, non-linear Schr\"{o}dinger equation of form
\begin{equation}
(\hat{\cal H} - i \hbar \partial_t) \psi({\bf r},t) = f(\psi({\bf r},t), \psi^{\dagger}({\bf r},t))
 \label{NLSE}
\end{equation}
where $f(\psi({\bf r},t), \psi^{\dagger}({\bf r},t))$ is some arbitrary function of the matter wave-function and its conjugate, whose origin is supposed to be gravitational (and so negligible for microscopic masses). Equations like this can be readily generalized to relativistic fields (see \cite{Kibble:1978vm,Kibble:1979jn} for details), where the non-linear term now depends on the expectation value of the field.

However, irrespective of the theory involved, equations like these run into severe difficulties. It was noted early on \cite{bialynicki1976nonlinear} that the rescaling of the wave-function during measurements required a specific logarithmic form for the potential, and it is hard to make theories of this kind consistent if one also assumes conventional ideas about quantum measurements. Further investigations of non-linear Schr\"{o}dinger equations by Weinberg \cite{Weinberg:1989cm,Weinberg:1989us} uncovered similar problems. The work of Weinberg, although it had nothing to do with gravitation, had two great virtues: it encapsulated many rather general features that such non-linear generalizations of Schrodinger's equation should possess, and at the same time produced a specific theory for non-linearities at the microscopic scale which was experimentally testable (and indeed it was falsified within a year \cite{bollinger1989test,chupp1990coherence,shull1980search}).

Further difficulties afflicting any non-linear quantum theory were also found by Polchinski and others \cite{Polchinski:1990py,gisin1989stochastic,gisin1990weinberg,wodkiewicz1990weinberg}; these included non-violation of Bell inequalities, and superluminal signal propagation. In particular, the Schr\"{o}dinger-Newton model admits an explicit protocol for superluminal signalling \cite{Bahrami:2014gwa}. We note, however, that such effects would be quite invisible in experiments done so far, if the source of the non-linearity is assumed to be gravitational.

(iii) It seems very difficult to derive a consistent theory including gravity, whether it be semiclassical or fully quantized, if one also tries to use the traditional quantum-mechanical framework of measurements, operators, observables, and states in Hilbert space. As noted already, this problem arises clearly in the non-linear Schr\"{o}dinger treatments, where normalization problems occur when measurement operations occur, and the usual Born rule is violated. It is certainly generic to any semiclassical treatment, which is inconsistent with the conventional wave function collapse - our example above showed this. Note that this problem is not obviated by assuming an ``epistemic'' interpretation for the wave-function, according to which $\psi$ only represents an observer's information about the world, rather than any ``real" state it may have. This is because the non-linearity of the time evolution of $\psi$ is sourced by $\braket{ T_{\mu\nu}}$, which cannot be observer dependent or undergo sudden ``wave function collapse"; for more discussion of this point, see remarks by Unruh \cite{unruh86} and Kibble \cite{isham1981quantum}.

In spite of all these problems, the semiclassical model can serve as a useful heuristic picture, and one can imagine attempting to test it in some limited sense, in which we simply accept the interpretation of $\ket{\psi}$ as a wavefunction and blindly apply the Born rule. A basic prediction of this type is the lack of entanglement production in the matter wave experiment as discussed in section \ref{classical-gravity}. Another early proposal, described by Carlip and Salzmann in 2006, was to look for a loss of coherence in matter-wave Talbot-Lau interferometry \cite{Salzman:2006rk,Carlip:2008zf,giulini2011gravitationally}, and more recently there have been proposals to look for distortions in the energy spectrum of optically trapped nanoparticles \cite{Grossardt:2015moa,Yang:2012mh,Helou:2017von}.

\subsection{Collapse models}
 \label{pd-models}

The initial formulation of quantum mechanics involved a fundamental split between the classical and quantum worlds, with the associated ``Copenhagen interpretation" providing the glue between the two. In an attempt to make sense of this, von Neumann in 1932 proposed the idea of a succession of entanglements between ever larger objects (the ``von Neumann chain") truncated by a probabilistic ``wave function collapse" into one particular state \cite{vonN32}. The state in \eqref{GRsup2} is a typical von Neumann state, and one supposes that one has the transition
\begin{equation}
\ket{\Psi_{in}} =  \ket{\Phi_o} \sum_k c_k \ket{\phi_k} \rightarrow \sum_k c_k \ket{\Phi_k; \phi_k^{\prime}}
 \label{vNmmt}
\end{equation}
where the initial ``system" state $\sum_k \ket{\phi_k}$ couples to the initial ``apparatus" state $\ket{\Phi_o}$ in such a way that these two systems entangle with a perfect correlation between the initial system states $\ket{\phi_k}$ and the apparatus states $\ket{\Phi_k}$, where we allow the final system states $\ket{\phi_k^{\prime}}$ to be different from the initial ones. This perfect correlation is the defining property of a perfect measurement operation (in the basis frame of the $\{ \phi_k \}$). However, to then engineer a definite final state with probability $P_k = |c_k|^2$, we require a ``collapse" of the superposition into just one of its branches, the $k$-th component. The lack of any theory whatsoever for how this collapse occurs is the famous measurement problem.

In what follows we only discuss ideas that have attempted to get around this problem using a gravitational mechanism. There are basically two of these, stochastic collapse models invoking a gravitational noise source, and Penrose's gravitational collapse model. In their simplest forms, these models also lead to a kind of semiclassical model of gravitationally-induced wavefunction collapse \cite{Diosi:2014ura,Penrose:1996cv}. Because this leads to non-linearity in the Schr\"{o}dinger equation, it is then clear that this idea will have problems of the kind just discussed \cite{Bahrami:2014gwa}. Thus any theory of this kind has to address problems of internal consistency.

\subsubsection{Stochastic collapse models}
Stochastic models of quantum mechanics have been advanced for a variety of reasons over a long period of time. In the non-relativistic limit these models have an equation of motion of general form
\begin{equation}
(\hat{\cal H} - i \hbar \partial_t) \psi({\bf r},t)  =  \xi(\braket{ \psi | F(\hat{O}_j(t)) | \psi}, t)
 \label{Sch-var}
\end{equation}
where $F$ is some arbitrary function, and the argument of the ``noise'' term $\xi(\langle F \rangle)$ involves a set of operators $\hat{O}_j(t)$ acting in the Hilbert space of the system. Work of this kind \cite{bassi2013models}, motivated by the quantum measurement problem, almost always deals with non-relativistic QM, and assumes the usual QM structure of state vectors, Hilbert space, and projective quantum measurements. The noise term may depend on $|\psi \rangle$, making (\ref{Sch-var}) non-linear. Note that in contrast to any uncertainty principle arguments \cite{karolyhazy1966gravitation}, this approach definitely leads to decoherence, and ``wave-function collapse'' caused precisely by the stochastic noise field.

Diosi \cite{diosi1986universal}, Ghirardi et al. \cite{ghirardi} and Pearle \cite{pearle1989combining} have discussed the possibility of a gravitational origin for this noise (here we will follow the treatment of Diosi).
In their models, the state of the matter system is taken to evolve according to a Schr\"{o}dinger equation with a ``noise" potential, superficially similar to the starting point of the Schr\"{o}dinger-Newton equation \eqref{scg-nr2}. One has
\be
i \hbar \partial_t \ket{\psi} = \left( H_{mat}  + \int d^3\mb{x} \ M(\mb{x}) \Phi(\mb{x}) \right) \ket{\psi},
\ee
where $M(\mb{x})$ is a mass density operator for the system, similarly to the mass density operator in section \ref{classical-gravity}. In the SN model, the potential $\Phi$ is a semiclassical quantity sourced by the state $\ket{\psi}$ itself. In the noise model, on the other hand, this potential is a stochastic quantity independent of $\ket{\psi}$, whose statistical correlations involve the gravitational propagator:
\be
\braket{ \Phi(\mb{x},t) } = 0, \ \ \ \braket{ \Phi(t_1,\mb{x}) \Phi(t_2,\mb{y}) }= G_N \frac{\delta(t_1-t_2)}{|\mb{x}-\mb{y}|},
\ee
where $\braket{ \cdots }$ represents an average over realizations of the noise. Performing this average, we obtain an evolution equation for the system density matrix:
\be
\partial_t \rho = -{i \over \hbar} [H_{mat},\rho] - \frac{G_N}{2} \int d^3\mb{x} d^3\mb{y} \frac{[M(\mb{x}),[M(\mb{y}),\rho]]}{|\mb{x}-\mb{y}|}.
 \label{diosiD}
\ee

This equation leads to decoherence in position space, i.e. decay of the off-diagonal density matrix elements. Our definition of the mass density operator in section \ref{classical-gravity} involves pointlike mass distributions. This causes the integral in \eqref{diosiD} to diverge! To remove this divergence, the massive objects were taken to have a uniform density. For the case of a spherically symmetric, uniform sphere of radius $R_0$, one gets a ``smeared" mass density operator
\be
M(\mb{x}) = \frac{m}{4 \pi R_0^3/3} \int d^3\mb{x}' \theta(R_0 - |\mb{x} - \mb{x}'|) \delta(\mb{x} - \hat{\mb{r}}),
 \label{massO}
\ee
where $\hat{\mb{r}}$ means the usual single-particle position operator, and $\theta$ is a step function. Then a position eigenstate $\ket{\mb{x}_0}$ of the particle is an eigenstate of this mass operator, with eigenvalue $M(\mb{x} | \mb{x}_0)$ given by
\be
M(\mb{x}) \ket{\mb{x}_0} = M(\mb{x} | \mb{x}_0) \ket{\mb{x}_0} = \frac{m}{4\pi R_0^3/3} \theta(R_0 - |\mb{x}-\mb{x}_0|) \ket{\mb{x}_0}.
\ee

With these assumptions in hand, one finds that the off-diagonal position-space density matrix elements decay exponentially in time
\be
\rho_{\mb{x}\mb{y}}(t) = \braket{ \mb{x} | \rho(t) | \mb{y}} \sim \rho_{\mb{x}\mb{y}}(0) e^{-\Gamma_{PD}(\mb{x},\mb{y}) t}
\ee
with rate
\be
\label{PDrate}
\Gamma_{PD}(\mb{x},\mb{y}) = \frac{G_N}{2 \hbar} \int d^3\mb{x'} d^3\mb{y'} \frac{ \left[ M(\mb{x}' | \mb{x}) - M(\mb{x}' | \mb{y}) \right] \left[ M(\mb{y}' | \mb{x}) - M(\mb{y}' | \mb{y}) \right]}{|\mb{x}'-\mb{y}'|}.
\ee
We see that superpositions between different positions $\mb{x} \neq \mb{y}$ will be damped in time at a rate proportional to the Newton constant.

The decoherence rate $\eqref{PDrate}$ has some simple features that can be read off directly. For one thing, if there is no superposition--i.e. $\mb{x} = \mb{y}$--we clearly have no decoherence, that is $\Gamma = 0$. Another feature is the behavior in the limit of a ``wide'' superposition $| \mb{x} - \mb{y} | \to \infty$. In \eqref{PDrate}, there are two types of terms, self-interactions involving only one branch of the wavefunction $M(\mb{x}' | \mb{x}) M(\mb{y}' | \mb{x})$, similarly for $\mb{y}$, and cross-interactions between the $\mb{x}$ and $\mb{y}$ branches $M(\mb{x}' | \mb{x}) M(\mb{x}' | \mb{y})$, similarly for $\mb{y}'$. The self-interaction terms comes with a positive sign while the cross-terms come with a minus and thus actually supress the decoherence. As we take the two branches to be well separated $|\mb{x} - \mb{y}| \to \infty$, these cross-terms vanish, and so the decoherence rate is maximized. This is quite reasonable; two widely-separated states are much easier for the background ``noise'' to distinguish, and thus this state should decohere faster, much as it would due to decoherence from random interactions with a thermal background (see eg. \cite{gallis1990environmental}). Indeed, precisely as in that case, the maximal decoherence rate limits to a constant--it does not grow indefinitely with the spatial separation of the wave function.

For example, consider our matter-wave experiment in figure \ref{cartoon-mw}. Assuming that the particles are well-separated with respect to our cutoff scale $\Delta x/R_0 \gg 1$, the decoherence rate \eqref{PDrate} can be easily estimated. The cross-terms simply vanish, and we are left with only the self-interaction terms, leaving
\be
\Gamma = \frac{G_N}{2 \hbar} \int \int d^3\mb{x'} d^3\mb{y'} \frac{ M(\mb{x}' | \mb{x}) M(\mb{y}' | \mb{x}) + M(\mb{x}' | \mb{y}) M(\mb{y}' | \mb{y}) }{|\mb{y}'-\mb{y}'|} \approx \frac{G_N m^2}{\hbar R_0}
\ee
where the approximation holds up to some dimensionless geometric factor of order one. Note that, by the argument in the previous paragraph, the dependence on the spatial separation $\Delta x$ between the beam arms has dropped out--the dominant part of the integral is the self-interaction terms. Observation of coherent oscillations in a matter-wave interferometry experiment then give lower bounds on this decay rate, in turn giving a bound on the free parameter $R_0$; for example, experiments using large organic molecules \cite{eibenberger2013matter} with $m \sim \SI{e5}{\amu}$ and $\Delta t \sim \SI{1}{\second}$ give $R_0 \gtrsim \SI{e-20}{\meter}$. One can do better with different types of interferometry; we refer the interested reader to \cite{bassi2017gravitational} for the state of the art.

Clearly there are several problems with such stochastic theories. A simple objection is that they are entirely non-relativistic. Worse, the prescription in \eqref{massO} for the mass operator is highly arbitrary; by varying the mass distribution one can vary the quantitative predictions of the theory over a very wide range. Finally, these models are {\it ad hoc}. They address a single problem, the measurement problem, in isolation; no general reason is given for introducing the stochastic fields, whose physical nature is not clarified, and whose effect on other physical phenomena is hardly discussed. Extra fields introduced in this way will likely conflict with other parts of physics; even if this is not the case, they will have testable effects on many other physical phenomena.

\subsubsection{Penrose collapse model} 

A quite different line of thought which leads in the simplest approximation to a very similar result to that above, was given by Penrose. The reasoning behind Penrose's approach can be seen by going back to the state $\ket{\Psi}$ in \eqref{GRsup}, in which the gravitational field is entangled with the matter field. Now let us consider what happens when we take the inner product $\langle \Psi | \Psi \rangle$, in which there will be interference terms between the two branches of the wavefunction. Since the state includes the quantum state of the gravitational field, we are faced with interference terms of the form\begin{equation}
\braket{ \Phi_L; g^{(L)}_{\mu \nu} | \Phi_R; g^{(R)}_{\mu \nu}}
 \label{GRint}
\end{equation}
between the two states of the matter and metric. How are we to evaluate this expression? In particular,  the two metrics $g^{(L)}$ and $g^{(R)}$ have in general different causal structures, and matter fields propagating on these background metrics would have different vacua, so it is not clear how to assign a meaning to this interference term.

Penrose has proposed that, in fact, there is an inherent difficulty in this problem precisely because of the differing causal structures related to the two metrics. He considered a special case where the two interfering states are stationary states, and found the difference in Newtonian gravitational potentials between the two metrics to be
\begin{equation}
\Delta E^g_{LR} \;=\;  4 \pi G \int d^3r \int d^3r' {\Delta M_{LR}({\bf r})\Delta M_{LR}({\bf r'}) \over |{\bf r} - {\bf r'}|} 
 \label{penrose}
\end{equation}
where $\Delta M_{LR}({\bf r}) = M_L({\bf r}) - M_R({\bf r})$, and the $M_j$ are the mass density distributions associated with each component of the wave function.

We notice the similarity in form of this result to that found for the stochastic decoherence rate, and we may analyze it in the same way as we did equation \eqref{PDrate}. However, as it stands $\Delta E^g_{LR}$ is an energy uncertainty, related to a time uncertainty $\Delta t_{LR}^g = \hbar/\Delta E^g_{LR}$. Penrose then makes the key step of identifying $\Delta t_{LR}^g$ as a decoherence time, to be viewed as an {\it intrinsic decoherence} existing in nature, implying a breakdown of quantum mechanics.

Given the radical difference between the underlying physical arguments leading to (\ref{penrose}) and those leading to (\ref{PDrate}), we feel that it is a mistake to treat the Penrose result as equivalent to the stochastic result, as is sometimes done. In particular, while Penrose's model is limited to the collapse of a massive superposition, the noise-based models are more general; it seems implausible that the noise would \emph{only} lead to collapse of massive superpositions. Nevertheless, at the level of the simple decoherence experiments considered in this paper, both models seem to give the same predictions.

Both collapse models suffer from the same imprecision in the definition of the mass density $M({\bf r})$. This point was made very clearly by Kleckner et al. \cite{kleckner2008creating}, who compared the results for $\Delta t_{LR}^g$ obtained by either (a) using the a Gaussian form for $M({\bf r})$ representing the ground state wave-packet, or (b) a density profile concentrated in a lattice of atomic nuclei, each of radius $10^{15}~$m; in the first case one finds $\Delta t_{LR}^g \sim 1~$sec, and in the second, $\Delta t_{LR}^g \sim 10^{-3}~$sec.

Testing either type of collapse theory proceeds most easily by preparing a single object in some kind of superposition state, and looking for it to decohere with the rates determined here. Experimental tests of the Penrose theory have been discussed, notably by Marshall et al. \cite{marshall2003towards} (see also \cite{kleckner2008creating}). They propose an interesting design in which a photon exists in a superposition of states located in two different optical cavities; in one of these cavities, the mirror is on an elastic spring, and is displaced when the photon is in the cavity, thereby entangling photon and mirror, and putting the mirror in a superposition of states at different locations. Original estimates for a mirror mass of $5 \times 10^{-12}~$kg and oscillation frequency $500~$Hz (giving a zero point spreading $\sim 10^{-13}~$m), show that experiments to look for gravitational decoherence here would be feasible - they have not yet been done.

\subsection{Correlated Worldline Theory}
 \label{cwl}

In sections \ref{scg} and \ref{pd-models}, we studied models in which gravity has been invoked as a source of non-linear quantum evolution. These models have been formulated in the language of canonical quantization and in a non-relativistic limit. In the Schr\"{o}dinger-Newton approach, the gravitational field is treated as a fundamentally classical degree of freedom; in the collapse models of Penrose, Diosi, et al., the actual dynamics of the gravitational field are left unspecified. In neither case was any attempt made to set up a real theory - the goal was to investigate the putative collapse process in isolation.

The difficulties of setting up a new theory have already been highlighted in our discussion of the efforts of Kibble and Weinberg, from which we learned that any semiclassical treatment (in which the gravitational field is not quantized like the other fields) leads to apparently insuperable problems, and that these are compounded if we try to keep the usual theoretical superstructure of measurements, observers, operators, etc., that characterize conventional quantum mechanics and quantum field theory.

Thus it seems reasonable to attempt a new theory involving wholesale reconstruction, rather than just tinkering with isolated bits of the theory, while at the same time keeping essential elements of quantum mechanics and general relativity. And - a key point - the new theory has to be internally consistent as well as being consistent with general physical principles and existing experiments.

An attempt at such a theory has recently been outlined in several papers\cite{stamp2012environmental,Stamp:2015vxa,barvinsky2018structure}, and given the name ``Correlated Worldline" (CWL) theory.  Although the formalism is rather complex in parts, the basic ideas are simple enough that we can give a brief outline here. A more detailed introduction to the rationale and assumptions behind the theory is also available \cite{Stamp:2015vxa}. From a purely theoretical point of view, one of the main attractions of the CWL theory is that it shows that a consistent theory of quantum gravity in which quantum mechanics is non-linearly modified is actually possible. However it also has experimentally testable implications, as we shall see.

The CWL theory is formulated in the language of path integrals, as this affords a very direct route to one thing we want to keep in the theory, which is the connection between action and quantum phase. To explain CWL theory in outline we first recall the standard path integral representation for the dynamics of a single particle, which propagates from position $\mb{x}$ at time $t$ to position $\mb{x}'$ at time $t'$ with amplitude
\be
U(\mb{x}',t';\mb{x},t) = \int_{\mb{q}(t)=\mb{x}}^{\mb{q}(t')=\mb{x}'} Dq \ e^{-i S[q]}.
\ee
Here $S[q]$ is the action functional and the integral is a sum over all paths $\mb{q}(t)$ with the specified boundary conditions. We note that this is just a linear sum, with one term for each path; this linear summation expresses the superposition principle in the path integral, and it makes $U(\mb{x}',t';\mb{x},t)$ unitary.

If we include the gravitational field dynamics into the theory, and look at a matter field with action $S_M$ instead of just a particle, this generalizes to
\be
\label{1paction}
U(\mb{x}',t',\bar{g}_{\mu\nu}';\mb{x},t,\bar{g}_{\mu\nu}) = \int_{x}^{x'} Dq \int_{\bar{g}_{\mu\nu}}^{\bar{g}_{\mu\nu}'} Dg_{\mu\nu} \ e^{-i S_{M}[q,g_{\mu\nu}] - i S_{G}[g_{\mu\nu}]}.
\ee
where we now also have a path integral over all configurations of the metric $g_{\mu\nu}(x)$ between metric configurations $\bar{g}_{\mu\nu}(x)$ and $\bar{g}_{\mu\nu}'(x)$. The action $S_{G}$ is the Einstein-Hilbert action; and again we note that this expression is still just a linear sum over configurations of the particle and metric. This action is the one from which we would ordinarily derive the effective field theory studied in section \ref{EFTsection}; it is the standard action for conventional quantum gravity. We note that there is, implicit in this path integral, some sort of UV cutoff to deal with UV divergences.

\begin{figure}[t!]
\centering
\includegraphics[scale=0.9]{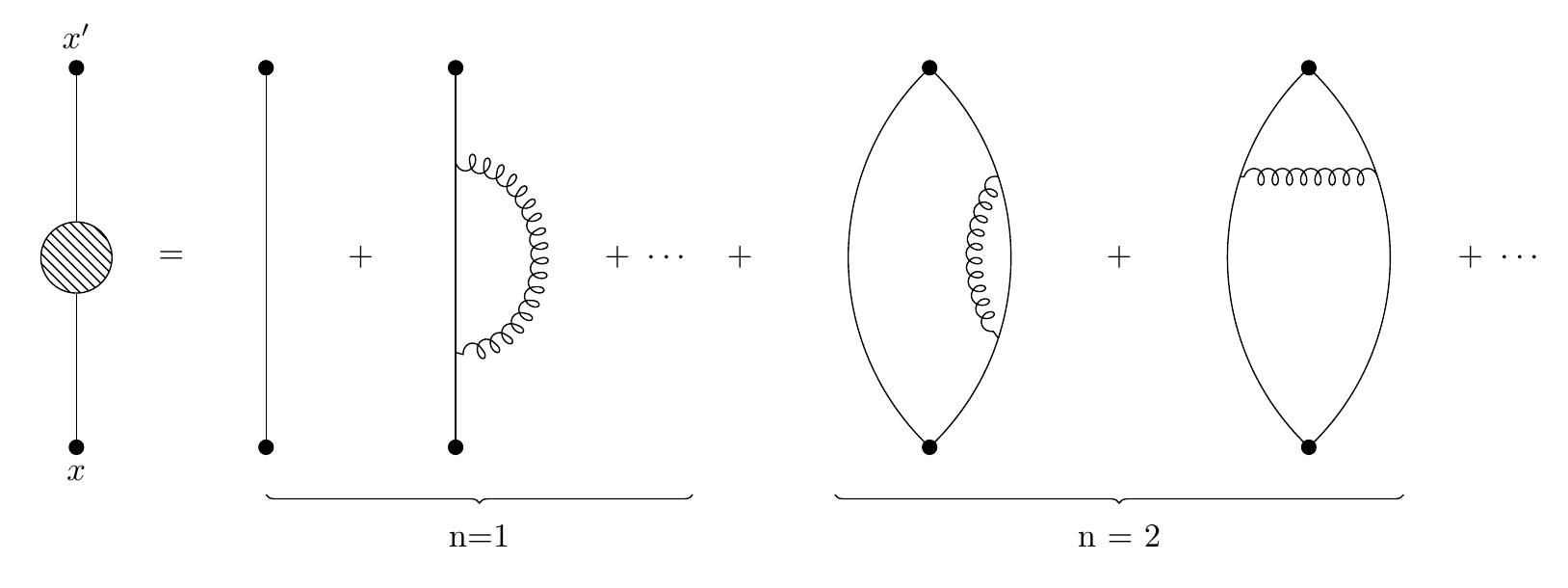}
\caption{Diagrammatic representation of the correlated worldline theory for a single-particle propagator. The solid black lines represent particle paths $q(t)$ between spacetime points $x$ and $x'$. The $n=1$ terms include the usual kinds of graviton loops. The $n=2$ terms, which correlate a pair of particle paths $q_1(t), q_2(t)$, have both the usual kinds of graviton loops (first diagram) as well as path-path correlations induced by gravity (second diagram). Ellipses represent higher-order terms in the graviton expansion and in CWL level $n \geq 3$.}
\label{cwl-fig}
\end{figure}

The proposal in \cite{stamp2012environmental,Stamp:2015vxa} is to now enlarge the theory by allowing for correlations between various paths $q,q'$ in the path integral. This automatically violates the superposition principle. It is then further argued that these correlations originate in gravity. This gives a special place to gravity in the theory, but we note that $g^{\mu\nu}(x)$ is still treated as a quantum field. By fairly lengthy arguments \cite{Stamp:2015vxa} one can start with the most general possible way of correlating the different paths, and then arrive at a specific form for gravitational correlations, using physical arguments based on the equivalence principle. Crucially, it turns out there are two different possible CWL theories \cite{barvinsky2018structure}, but one of them fails certain consistency tests. In what follows we explain the basic idea, and discuss how the CWL predictions for experiment can be made, and how they differ from conventional quantum mechanics. These formulations are:

(i) {\bf Summation form:} In this form, we sum over all possible correlations between different paths in the propagator - see ref. \cite{Stamp:2015vxa} for a derivation. The CWL propagator is then an infinite sum $K = \sum_k K_n$, of form
\be
K_{CWL}(\mb{x}',t',g';\mb{x},t,g) = \sum_{n=1}^{\infty}  \int_{x}^{x'} \prod_{i=1}^{n} Dq_{i} \int_{g}^{g'} Dg \ e^{-i \sum_i S_{M}[q_i,g] - i S_{G}[g]}.
\ee
Note that $K$ is not in general unitary. The key point here is that the path integral over the metric configurations correlates all the paths. The $n=1$ term is equivalent to \eqref{1paction}; it is just standard quantum gravity, and so reproduces the predictions of the effective field theory picture in section \ref{EFTsection}. The next $n=2$ term is
\be
\label{CWL2}
K_{2}(\mb{x}',t',g';\mb{x},t,g) = \int_{x}^{x'}  Dq_1 Dq_2 \int_{g}^{g'} Dg \ e^{-i S_{mat}[q_1,g] - S_{mat}[q_2,g] - i S_{EH}[g]}.
\ee
in which each path $q_1$ in the path integral couples to every other path $q_2$ via the gravitational interaction. The higher order terms $n=3, 4, \ldots$ allow more complex correlations to develop. Figure \ref{cwl-fig} shows a diagrammatic representation of calculations of the propagator in an expansion in gravitons.

As long as we are working with weak gravitational sources, for example particles in a laboratory, we can ignore CWL correlations beyond $n=2$, since these will be suppressed by powers of the Planck mass. The $n=2$ term comes in at the same order in perturbation theory in the Planck mass as the $n=1$ term: both terms come from single-graviton exchange diagrams (see fig. \ref{cwl-fig}), in the $n=1$ term from graviton exchange between two different particles, whereas in the $n=2$ term from graviton exchange between different copies of the \emph{same} particle. This then leads to an interesting pattern of correlations, in which we begin with the standard correlations from the $n=1$ term and progressively destroy them through the $n \geq 2$ terms.

There is a very concise way of formulating the summed CWL theory, which is to write a generating functional (the analogue of the partition function in statistical mechanics) from which all propagators, correlation functions, etc., can be derived. For the summed CWL theory, this takes the form \cite{Stamp:2015vxa}
\begin{equation}
\mathbb{Q}[J] \;=\;  \oint {\cal D}g \; e^{{i \over \hbar} S_G[g]} \sum_{n=1}^{\infty} {1 \over n!}\;Q_n[g,J]  \;\;\;\;\; (summed) \;\;\;
 \label{Q-sch1}
\end{equation}
in which $Q_n[g,J]$ is that contribution to the sum coming from an $n$-tuple of paths, and $J(x)$ is an external "source" current coupled to the matter field. One then obtains propagators and correlators automatically, just by functionally differentiating $\mathbb{Q}[J]$ with respect to $J(x)$; for details see refs. \cite{Stamp:2015vxa,barvinsky2018structure}.

However, it turns out there is a fly in the ointment - a recent more detailed investigation \cite{barvinsky2018structure} has found that the summed version of CWL theory has an inconsistency in the semiclassical limit - thus it has to be discounted, although many useful lessons can be learnt from it.

(ii) {\bf Product form:} It turns out that one {\it can} derive a CWL theory which passes all consistency tests: it has sensible semiclassical and perturbative expansions, and satisfies all Ward and Noether identities. We can compare this with summed CWL very simply by writing a new generating functional
 \begin{eqnarray}
    &&\mathbb{Q}[J]=\prod\limits_{n=1}^\infty \tilde{Q}_n[\,J\,]  \;\;\;\;\;\; (product) \;\;\; \\
    &&\tilde{Q}_n[\,J\,]=
    \oint {\cal D}g \; e^{{i \over \hbar} S_G[g]} \,\left({\cal Z}[g,
    J/c_n]\right)^n     \label{bbQ-J01}
    \end{eqnarray}
where ${\cal Z}$ is just the particle generating functional in conventional field theory (with a regulator $c_n$ that we do not discuss here);  we are now summing over logarithms, and $\ln \mathbb{Q}[J]$ is now used to generate correlation functions.

At first glance all one seems to have done here is substitute a logarithm of a product in place of a sum - why is there any difference? But it turns out that the correlations between paths that are generated come with a different weighting, and this cures the problems in the original summation form. We still have correlations between paths and so the superposition principle is still violated. To see how this works in the same kind of perturbative expansion that we just described for summed CWL theory is a little more difficult in the product version. If one evaluates the lowest order graphs, one still sees that correlations develop between different paths. However they now appear as a kind of gravitational attraction of the different graphs towards the vacuum energy generated by each graph. See \cite{barvinsky2018structure} for more on the technical details.

The principal change in the physics caused by the correlations in CWL theory is what is called "path bunching"; the attractive gravitational interaction between the paths causes them to start clustering together once the mass of the propagating object is large enough \cite{Stamp:2015vxa,barvinsky2018structure}. This is quite different from standard quantum mechanics, where one sums over independent paths, with the differences between the phases of the different paths giving the usual interference phenomena. In CWL theory, path bunching prevents paths from separating widely once the mass exceeds $m_{CWL}$, at which point double-slit interference no longer becomes possible - the paths cannot separate to encompass both slits, and the mass starts to behave classically.

\begin{figure}[t!]
\centering
\includegraphics[scale=0.35]{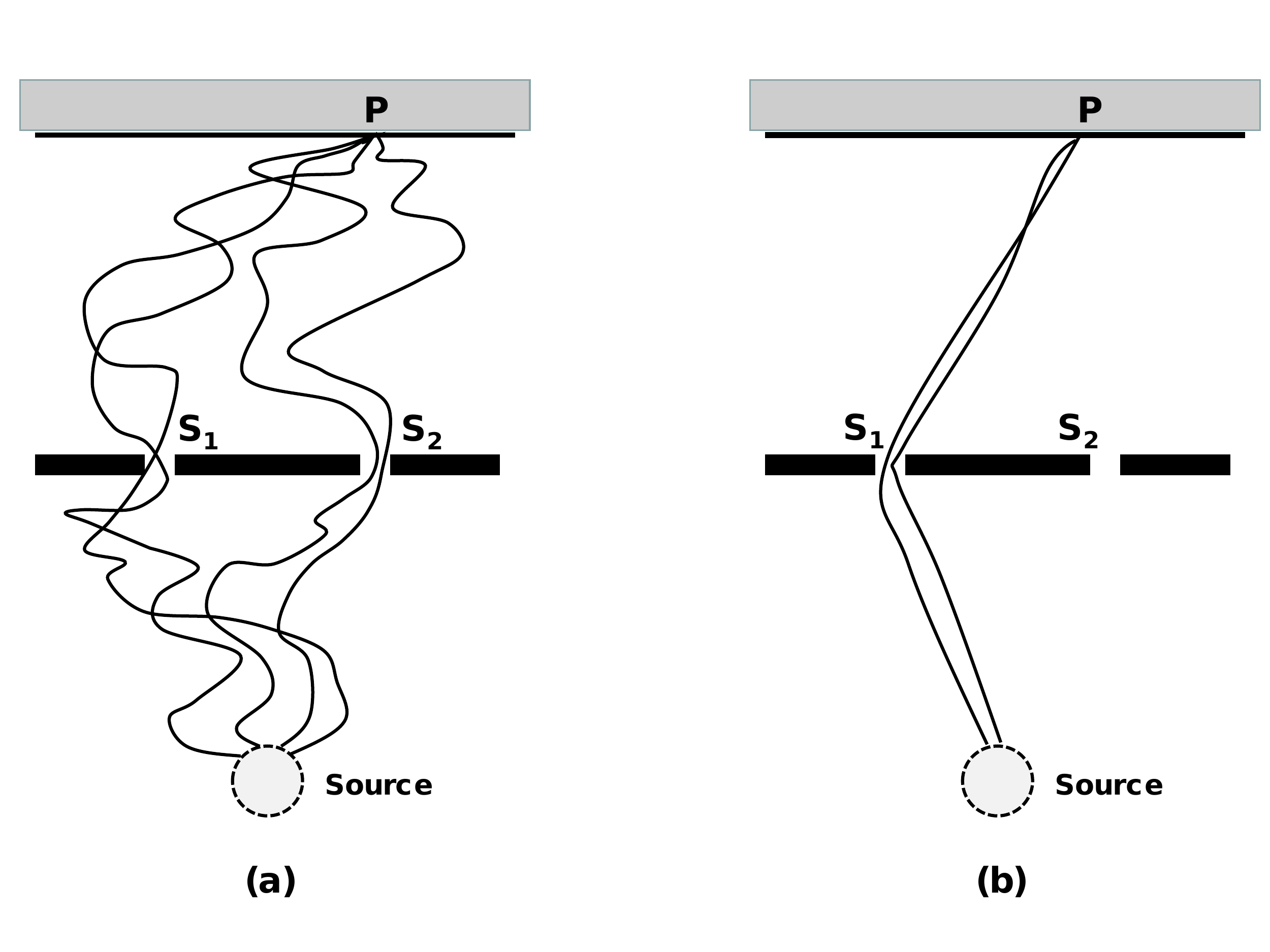}
\caption{The effect of path bunching on a double-slit experiment. In (a) we see typical paths contributing to the amplitude for a microscopic particle to pass from a source, through a pair of slits $S_1$ and $S_2$, to a point $P$ on a screen - conventional quantum mechanics is obeyed. In (b) we see one class of paths contributing to the same amplitude when the particle mass exceeds $m_{CWL}$; the paths are all strongly attracted to each other, and all the paths pass through slit $S_1$. There is another disjoint class of paths passing through $S_2$, which do not interfere with those passing through $S_1$.}
\label{cwl-fig-2}
\end{figure}

The physics of path bunching is illustrated in figure \ref{cwl-fig-2}, which contrasts the kinds of path contributing in a double-slit experiment to the amplitude to propagate to a given point on a screen in conventional quantum mechanics, with that prevailing in CWL theory for a sufficiently large mass. In the latter case, a set of paths passing through one slit cannot be accompanied by paths traveling through the other, since the latter paths are too strongly attracted by the former.

Note there is no decoherence involved here in the ``quantum-classical transition'' between purely quantum behavour and the classical behavior for large masses; no external environments or noise are involved. Another key feature of the theory is that if a microscopic system interacts with a massive measuring system, so that they become entangled, then path bunching occurs for the ``combined system + apparatus'', and the pair of them now show classical behavior. As a general result the crossover between quantum and classical dynamics happens very quickly as one increases the mass of the objects involved \cite{barvinsky2018structure}; unless the mass happens to be very close to $m_{CWL}$, one sees either quantum dynamics or classical dynamics.

{\bf Experiments}: The CWL theory has no adjustable parameters, and so testable predictions can in principle be made. However we emphasize that to date there is still no reliable result for the crossover mass $m_{CWL}$ in the product version of CWL theory. From calculations in summed CWL theory, it is known that the crossover mass depends on the detailed composition of the object - its shape and density, and even its phonon spectrum - but these are all readily determined in advance. It is expected to be rather large for typical solid bodies. In the summed CWL theory one finds that the crossover mass is typically about $10^{-9}$ kg, not much less than the Planck scale of $2.2 \times 10^{-8}$ kg. The crossover mass in the product CWL theory is likely to be somewhat larger - a detailed evaluation has yet to be given.

One class of experiments involves Talbot-Lau interferometry, where we only have one particle to deal with. In this case, the prediction would be a disappearance of the interference pattern above a certain mass scale. We refer the reader to \cite{stamp2012environmental,Stamp:2015vxa} for some more detailed discussion of this type of experiment. Another experiment may be more likely to succeed in the near term. The set-up of Marshall et al. \cite{marshall2003towards} mentioned earlier (or some variant of it), involving superpositions of mirrors on springs, should allow spatial superpositions of much more massive objects than a Talbot-Lau set-up. If superpositions involving masses $\sim 10^{-9}$ kg can be formed, this would offer an ideal route towards testing CWL theory.

\section{Conclusions}
\label{conclusions}

Rapid advances in the quantum control of the state of meso-to-macroscopic systems, combined with advances in quantum sensing of tiny forces, have opened the possibility of the first direct observation of quantum aspects of the gravitational interaction.

We have emphasized that the usual tenets of quantum field theory are in perfect harmony with general relativity at the terrestrial scales of interest here. Gravity can be viewed as an effective quantum field theory, which breaks down at extremely high energies but is perfectly capable of making predictions for tabletop experiments. On the other hand, the experimental situation is still wide open: gravity may well be quantum, and we have endeavored to provide a representative set of models of ``classical'' gravity, and their predictions in some prototypical low-energy experiments. In this paper, we have emphasized that a number of these models--including the quantum field theory picture--will be testable with near-future technology.

This line of thinking presents many opportunities for both experimentalists and theorists. Experimentally, it goes without saying that control over larger masses, superposed on larger spatial scales, and with longer coherence times will be invaluable. More detailed characterizations of non-gravitational decoherence channels are critically needed: distinguishing the tiny gravitational signatures of interest from other, non-gravitational backgrounds will be crucial. On the theoretical side, further generation and characterization of models of low-energy gravity would be of significant interest and utility. In particular, the precise behavior of ``emergent'' gravity models in the kinds of scenarios envisioned here is an area ripe for exploration.

After nearly a century, it appears that the dream of observing the quantum nature of gravity may finally be near at hand. That this may occur on a tabletop rather than near a black hole or in a high-energy collider is unexpected and deeply exciting. We have endeavored here to help set the stage for this exciting and emerging paradigm, and look forward to rapid development of these ideas in the near future.

\section*{Acknowledgements}

Thanks to Markus Aspelmeyer, Andrei Barvinsky, Jacques Distler, John Donoghue, Jack Harris, Nobuyuki Matsumoto, Eugene Polzik, Jon Pratt, Tom Purdy, Jess Riedel, Stephan Schlamminger, Julian Stirling, Clive Speake, Bill Unruh, and Denis Vion for discussions on various aspects of this work, and to Joel Meyers for critical reading of an early version of this paper. D. C. would also like to thank the theory group at Fermilab for their hospitality while a portion of this work was completed. The work of P. C. E. S. is supported by NSERC in Canada and the Templeton Foundation; D. C. and J. M. T. are supported by the NSF-funded Joint Quantum Institute and by NIST.

\bibliographystyle{utphys-dan} 
\bibliography{ms-grav-refs}

\end{document}